\documentclass{aa}

\newcommand{\teff}{$T_\mathrm{eff}$}
\newcommand{\logg}{$\log{g}$}
\newcommand{\feh}{[Fe/H]}

\usepackage{graphicx}
\usepackage{txfonts}
\usepackage{natbib}
\usepackage{amsmath}
\usepackage{xspace}
\usepackage{txfonts}
\usepackage{txfonts}
\usepackage{array}
\usepackage{color}
\usepackage{url}
\usepackage{enumitem}
\usepackage{bm}
\usepackage{xspace}
\usepackage{siunitx}
\usepackage[dvipsnames]{xcolor}
\usepackage[colorlinks = true,
            linkcolor = blue,
            urlcolor  = blue,
            citecolor = blue,
            anchorcolor = blue]{hyperref}
\usepackage{threeparttablex} 
\usepackage{longtable}
\usepackage{aas_macros}
\usepackage{siunitx}
\usepackage{nicefrac}

\begin{document}

   \title{Ariel stellar characterisation: I-- homogeneous stellar parameters of 187 FGK planet host stars}

   \subtitle{Description and validation of the method}
\titlerunning{Ariel homogeneous stellar parameters}
\authorrunning{Magrini et al.}
   \author{
          L. Magrini\inst{\ref{oaa}}, 
          C. Danielski\inst{\ref{iaa},\ref{iap}}, 
          D. Bossini\inst{\ref{porto}},    
          M. Rainer\inst{\ref{oaa},\ref{oabr}},
          D. Turrini\inst{\ref{oato}}, 
%
        S. Benatti\inst{\ref{oapa}},
        A. Brucalassi\inst{\ref{oaa}}, 
        M. Tsantaki\inst{\ref{oaa}}, 
        E. Delgado Mena\inst{\ref{porto},\ref{porto2}}, 
        N. Sanna\inst{\ref{oaa}}, 
        K. Biazzo\inst{\ref{oarm}},
        T. L. Campante\inst{\ref{porto},\ref{porto2}}, 
%
        M. Van der Swaelmen\inst{\ref{oaa}},        
        S. G. Sousa\inst{\ref{porto}}, 
        K. G. He{\l}miniak \inst{\ref{torun}},
        A. W. Neitzel \inst{\ref{porto},\ref{porto2}},
        V. Adibekyan\inst{\ref{porto}}, 
        G. Bruno\inst{\ref{oact}}\and
        G. Casali\inst{\ref{difa},\ref{oas}}
          }

\institute{
INAF - Osservatorio Astrofisico di Arcetri, Largo E. Fermi 5, 50125, Firenze, Italy \email{laura.magrini@inaf.it} \label{oaa} 
\and
Instituto de Astrofísica de Andalucía, CSIC, Glorieta de la Astronomía, E-18080, Granada, Spain\label{iaa}
\and
Sorbonne Universit\'es, UPMC Universit\'e Paris 6 et CNRS, 
UMR 7095, Institut d'Astrophysique de Paris, 98 bis bd Arago,
75014 Paris, France\label{iap}
\and
Instituto de Astrofísica e Ciências do Espaço, Universidade do Porto, CAUP, Rua das Estrelas, 4150-762 Porto, Portugal\label{porto}
\and
INAF - Osservatorio Astronomico di Brera, Via E. Bianchi 46, 23807 Merate (LC), Italy\label{oabr}
\and
INAF - Osservatorio Astrofisico di Torino, via Osservatorio 20, 10025, Pino Torinese\label{oato}
\and
INAF - Astronomical Observatory of Palermo, Piazza del Parlamento 1, 90134, Palermo (Italy)\label{oapa}
\and
Departamento de F\'{\i}sica e Astronomia, Faculdade de Ci\^{e}ncias da Universidade do Porto, Rua do Campo Alegre, s/n, 4169-007 Porto, Portugal\label{porto2}
\and 
INAF - Osservatorio Astronomico di Roma, Via Frascati 33, I-00040 Monte Porzio Catone (Roma), Italy \label{oarm}
\and
Nicolaus Copernicus Astronomical Center, Polish Academy of Sciences, ul. Rabia\'{n}ska 8, 87-100 Toru\'{n}, Poland\label{torun}
\and
INAF - Catania Astrophysical Observatory, Via Santa Sofia, 78, I-95123 Catania, Italy\label{oact}
\and
Dipartimento di Fisica e Astronomia, Università degli Studi di Bologna, Via Gobetti 93/2, I-40129 Bologna, Italy\label{difa}
\and
INAF - Osservatorio di Astrofisica e Scienza dello Spazio di Bologna, via Gobetti 93/3, 40129, Bologna, Italy\label{oas} 
}

   \date{}

 
  \abstract
   {In 2020 the European Space Agency selected Ariel as the next mission to join the space fleet of observatories to study planets outside our Solar System. Ariel will be devoted to the characterisation of a thousand planetary atmospheres, for understanding what exoplanets are made of, how they formed and how they evolve. To achieve the last two goals all planets need to be studied within the context of their own host stars, which in turn have to be analysed with the same technique, in a uniform way.} 
   {We present the spectro-photometric method we have developed to infer the atmospheric parameters of the known host stars in the Tier 1 of the Ariel Reference Sample. }
   {Our method is based on an iterative approach, which combines spectral analysis, the determination of the surface gravity from {\em Gaia} data, and the determination of stellar masses from isochrone fitting. We validated our approach with the analysis of a control sample, composed by members of three open clusters with well-known ages and metallicities. }
   {We measured effective temperature, \teff, surface gravity, \logg\/, and the metallicity, \feh\/, of 187 F-G-K stars within the Ariel Reference Sample. We presented the general properties of the sample, including their kinematics which allows us to separate them between thin and thick disc populations.    }
   {A homogeneous determination of the parameters of the host stars is fundamental in the study of the stars themselves and their planetary systems. Our analysis systematically improves agreement with theoretical models and decreases uncertainties in the mass estimate (from 0.21$\pm$0.30 to 0.10$\pm$0.02 M$_{\odot}$), providing useful data for the Ariel consortium and the astronomical community at large. }


   \maketitle
%

\section{Introduction}

In 2029 the ESA Ariel space mission \citep{Tinetti2018}  will join the James Webb Space Telescope in L2 with the goal of shedding light on the formation and evolution of exoplanets through the observation of a large sample of atmospheres. 
For maximising the mission scientific return, the Ariel Consortium has developed a three-tiered observing approach that builds from a large population study (Tier 1), to the increasingly more detailed characterisation of selected high-interest objects (Tier 3), all by taking full advantage of the mission unique characteristics \citep{tinetti21}.

Tier 1, known as the ``Reconnaissance Survey'', corresponds to the complete list of $\sim$1000 exoplanets that Ariel will observe between 0.5 -- 7.8 $\mu$m by means of photometry in the visible bands,  and low-resolution spectroscopy beyond 1.1 $\mu$m.
The atmospheric characterisation of the complete sample will allow us to know the fraction of planets that are covered by clouds, and the fraction of small planets that have still retained H/He \citep{Turrini2018, Tinetti2018,Edwards2019}. Furthermore, it will enable us to constrain and/or remove degeneracies in the interpretation of the mass-radius diagram, and to classify the planetary population by building colour-colour and colour-magnitude diagrams.
Yet, the observational methodology of Ariel is based on differential spectroscopy that measures planetary signals of 10 -- 50 ppm relative to the star, which entails the scientific importance of precisely characterising all host stars' properties beforehand.
Besides, it is likewise essential to establish a robust and bias-free stellar reference frame on which to build our knowledge on exo-atmospheres and planetary formation processes \citep{Turrini2021}. This translates into using the same uniform methodology for the analysis of all Tier 1 stars, and into making sure that the full set of parameters for each star is self-consistent \citep{Danielski2021}. Generally, the inconsistency of stellar parameters found in the literature did not turn out to be a fertile ground for robust statistical studies, which instead proved successful when homogeneous parameters were adopted (e.g., \citealt{Hartman2010,Adibekyan2015,Adibekyan2021,Biazzo2022}). 

To make a meaningful selection of the definitive Tier 1 targets, to bolster the possibility of finding hidden trends and correlations between stellar and planetary properties when data will flow in, and hence to optimise the science return of the mission, the ``Stellar Characterisation'' working group of the Ariel Consortium has committed to homogeneously characterise all stellar hosts included in the Ariel Reference Sample i.e., the list of potential targets to be observed by Ariel after its launch \citep{Zingales2018}. The Reference Sample employed for this study has been presented by \citet{Edwards2019} and it includes a total of 487 known planet hosts, plus a list of foreseen targets built up through the Ariel radiometric model with predicted Transiting Exoplanet Survey Satellite (TESS, \citealt{Ricker2014}) discoveries, for a total of $\sim$1000 potential targets with spectral types F-G-K and M (typically brighter than K\,=\,11 mag). The planets in the sample span different sizes from Jupiter to Earth-like planets, equilibrium temperatures from 500 K to 2500 K, and bulk densities from  0.10 to 10.0 g~cm$^{-3}$. Alongside with the ongoing new planetary discoveries, the Reference Sample can be reviewed to include more optimal targets for Ariel science as to fill the $\sim$1000 targets Tier 1. A very recent update of the Reference Sample has been presented by \cite{EdwardsTinetti2021} with the addition of a number of already confirmed TESS planets.

In this paper, we will expand on the study by \cite{Brucalassi2021} (hereafter B21), presenting the conclusive description and validation of our new iterative spectro-photometric method that we will be using for determining the effective temperature, \teff, surface gravity, \logg\/, and metallicity, \feh\/, of 187 F-G-K stars in the Ariel Reference Sample. 
Our work relies on both newly obtained and  publicly available high-resolution, high signal-to-noise ratio (S/N) data that we obtained through an ongoing ground-based monitoring campaign in both hemispheres. Our results will be included in an incremental way, along with the gathering of new spectroscopic data over the years,  into a much required homogeneous and self-coherent catalogue spanning many host-star parameters (i.e., atmospheric parameters, elemental abundances, activity indicators, ages, masses, radii) which will be made publicly available\footnote{The catalogue is currently available under request at \url{scwgariel@gmail.com}} 

The paper is structured as follows: we present in Sec. \ref{sec:samples} the sample of high-resolution spectroscopic data used for our analysis. We describe in Sec. \ref{sec:method} the method we have developed, together with details on the stellar mass determination in Sec. \ref{sec:mass}. In Sec. \ref{sec:validation}, we apply the method to three stellar clusters, as validation. We present our results in Sec. \ref{sec:results}, which we go on to discuss in Sec. \ref{sec:relations} in terms of their relation with their planetary companions' features. We conclude in Sec. \ref{sec:conclusions}.

\section{The samples}
\label{sec:samples}
In the present paper, we focus on Ariel F-G-K stars in the main sequence. Future works will be devoted to analyse other sets of stars e.g., with temperatures higher than F spectral-type and fast rotators (Paper II) and  M dwarfs (Paper III).

In what follows, we discuss: {\em i)} two samples of stars whose spectra are present in the public telescope archives, and {\em ii)} three samples with new observations we obtained in the past two years with the VLT, TNG and LBT telescopes. 
For the targets observable from the Southern hemisphere, we used the spectrograph UVES at VLT \citep{dekker00} for its combination of large mirror aperture and high spectral resolution. For the Northern targets, we used HARPS-N. at TNG \citep{Cosentino2012} for the brightest targets (V<12.5), complementing it with PEPSI at LBT \citep{Strassmeier03} for fainter targets.

\subsection{The archival sample of \citet{Brucalassi2021}}

The sample presented in B21 was produced by cross-matching the Ariel Reference Sample with the spectra available in the 2020 version of SWEET-Cat (Stars With ExoplanETs Catalogue\footnote{ \url{http://sweetcat.iastro.pt/}}), and by selecting all targets in common with parameters' source ``flag=1'', which
are the stars analysed homogeneously \citep[see][]{Santos2013,Sousa2021}.
The aim of B21 was to re-determine the stellar parameters of the Ariel-SWEET-Cat cross-matched sub-sample using two
different methods \citep[FAMA, Fast Automatic MOOG Analysis,][]{Magrini2013}, based on the equivalent width, and FASMA \citep[Fast Analysis of Spectra Made Automatically,][]{Tsantaki2018}, based on spectral synthesis, and to compare them with the original results of SWEET-Cat. 
The comparison among the three different methods identified some regions in the parameter space in which the spectral analysis produced less reliable results, especially for the determination of surface gravity. These regions are the upper and lower main sequence.
B21 suggested the  use of spectro-photometric gravity to improve the determination of stellar parameters. In B21, a first test was proposed comparing the stellar parameters, \teff\ from the spectral analysis and the surface gravity obtained from {\em Gaia} {\sc dr2} photometry and parallaxes, with a set of {\sc PARSEC} isochrones \citep{Bressan2012}. This first test already showed a considerable improvement over the parameters derived spectroscopically only, suggesting the method we describe in detail in the present paper.

We include in the present work the sample of 126 stars presented in B21. These are F-G-K stars with magnitudes in the range 5$<$V(mag)$<$16 and  with high-resolution spectra with S/N between $\sim$50 and $\sim$800.
Details on the archival spectra for this sample can be found in B21. 
Here we re-analyse them in a consistent way together with the other samples.

\subsection{The HARPS archival sample}
This sample was produced by cross-matching the Ariel Reference Sample with the spectra available in the HARPS-ESO Archive. The spectra have a very high spectral resolution (R=115,000) and they cover a wide wavelength range (380.0-690.0~nm). The initial sample was composed by 52 stars. After combining all the individual exposures for a given star, in some cases, the S/N was not good enough to derive parameters and in other cases the spectral type was not within the range we consider here. The sample of stars for which we do not deliver parameters includes some late K or M dwarf stars (GJ 1132, GJ 3470, GJ 9827, Wolf-503), some hot or rotating stars (HATS-24, HATS-41, HATS-64, HATS-65, WASP-80, HD 106315, WASP-121, NGTS-2), some suspected red giant stars (K2-3, K2-39, K2-132, K2-266, KELT-11), and  K2-287, a solar-type star. 
For other two stars, their spectroscopic metallicity was out of the isochrone grid and thus their mass and the final parameters could not be derived (HATS-60\footnote{For HATS-60 we provide stellar parameters from new acquired UVES spectra} and HD 89345). 
We present parameters for a final sample of 33 stars. Three of them are in common with new observations, although these spectra have a lower S/N. We keep these three stars for a consistency check. 

\subsection{The UVES sample}
We had two successful proposals (105.20P2.001 and 106.21QS.001, both with Principal Investigator (PI) C. Danielski) submitted to UVES at VLT: despite the pandemic delaying the observations we were able to collect data for 23 stars.
We used UVES with the following setup: Image Slicer \# 1 (R=60,000), central wavelength in the blue 390 nm, and in the red 580 nm. Our goal was to reach signal-to-noise S/N>120 at 600 nm, combining different exposures in order to minimise the cosmic rays' effect.\\
The spectra were reduced with the standard UVES pipeline and then combined in order to have two spectra (one for each full wavelength range) per target. For some of the P106 spectra the standard reduction was unsatisfactory due to a combination of low signal and a possible over-correction of the sky that resulted in the presence of a large number of spikes. We reduced these spectra again optimising the options of the pipeline, and we were then able to recover them.
We present parameters for a sample of 20 stars. Three stars were excluded from the present work since two of them are late-K or M dwarfs (K2-21 and HATS-22) and one is a hot rotating star (HATS-41). 

\subsection{The TNG sample}
The high resolution (R=115,000) of HARPS-N is well suited for our study. However, the smaller collecting area compared to the VLT (3.6 m versus 8 m) limited us to the brightest stars of our sample.
We completed successfully two proposals with HARPS-N (AOT41\_TAC25 -- PI S. Benatti, and AOT42\_TAC20 -- PI M. Rainer), and we collected data for 9 stars. As for the UVES data, we split the observations in shorter exposures to reduce the impact of the cosmic rays. The spectra were fully reduced by the HARPS-N DRS \citep{Cosentino2014}, and then we combined them in order to obtain a single high-S/N spectrum for each star.
Here, we present stellar parameters for 7 stars, excluding the two fast rotating stars (HAT-P-57 and KELT-1).

\subsection{The LBT sample}
For the fainter Northern targets, we submitted a proposal to PEPSI at LBT\footnote{\url{https://pepsi.aip.de}} (2021\_2022\_25 -- PI M. Rainer), selecting the lowest resolution mode (300 $\mu$m fiber, R=50,000). For each star we cycled on the cross-dispersers setup combination CD1+CD4, CD2+CD5, CD3+CD6 in order to cover the whole available wavelength range (393 -- 914 nm).\\
The observations are still ongoing, but we have already collected spectra for 6 stars. The spectra were reduced with the SDS4PEPSI package \citep{Strassmeier2018} by the PEPSI team. This resulted in a single, combined full-wavelength range spectrum for each star.
Here, we present stellar parameters for 4 stars, excluding a fast rotating star (Kepler-1517) and a late K dwarf (K2-77). 

\section{The method}
\label{sec:method}
Since about two decades ago, the importance of a homogeneous determination of the parameters of planetary host stars has been recognised \citep[see, e.g.,][]{fv05}. However, spectroscopic analysis can lead to some degeneracy between temperature and gravity, appreciable when the results are compared with the expected parameters from theoretical stellar models. 
Several attempts have been made to use external constraints \citep[e.g][]{Valenti09, Torres2012, Mortier2013} or to improve the linelist by including a large number of {\sc Fe~II} lines \citep[e.g][]{brewer15} to better constraint the surface gravity. 
The advent of the {\em Gaia} mission \citep{Gaia2018}, providing distances and luminosities of billions of stars, allowed us to provide further constraints to the determination of the surface gravity than has occurred in the past. 
In this framework, we developed an iterative spectro-photometric method to determine the stellar parameters,  which is based on the use of high-S/N and high-resolution spectra together with photometric and astrometric data from {\em Gaia} and from ground-based  surveys, as 2MASS \citep{Skrutskie06}. 
Our spectral analysis is based on the use of absorption line equivalent widths (EWs). We normalised the spectra with a global polynomial fit and measured the EWs with {\sc DAOSPEC} \citep{Stetson08}, through the wrapper {\sc DOOP} \citep{Cantat-Guadin2014}. We used the line list developed for the {\em Gaia}-ESO survey \citep{gilmore12, randich13} and presented in \citet{heiter21}. We adopted  the radiative transfer code {\sc MOOG} \citep{Sneden1973}, automatised with the wrapper FAMA \citep{Magrini2013}, to compute stellar parameters. 
We used a wide grid of MARCS model atmospheres \citep{Gustafsson08}, in both the spherical (for $\log{g}$ <3.5) and plane-parallel ($\log{g}\geq$3.5) configurations. 

To obtain the stellar parameters (effective temperature, \teff\/, surface gravity, $\log{g}$, metallicity, [Fe/H], and microturbulent velocity, $\xi$), we derived the iron abundances from the absorption lines of  Fe~{\sc i} and Fe~{\sc ii} lines, A(Fe~{\sc i}) and A(Fe~{\sc ii})\footnote{A(X)=12+log$_{10}$(N(X)/N(H)), where N(X)/N(H) is the abundance by number with respect to H of the element X.}. We used the iron abundances to establish: the excitation equilibrium from which we can infer the effective temperature, \teff\/; the ionisation balance to derive the surface gravity, $\log{g}$; and the equilibrium between A(Fe~{\sc i}) and the reduced equivalent width log(EW/$\lambda$) to derive the microturbulent velocity, $\xi$ \citep[see, e.g.,][]{Mucciarelli11}.
The stellar parameters  obtained in this way are, thus, the initial values of our process.

To derive our final set of stellar parameters, we have followed the steps (Runs), from Run~0 -$R.0$- to Run~2 -$R.2$ described below: 
\begin{itemize}
    \item[{\em R.0}] we started with a {\em pure} spectral analysis, deriving the first set of stellar parameters ($T_{\rm eff}0$, $\log{g}_0$, [Fe/H]$_0$, $\xi_0$) of our sample stars from the excitation equilibrium, ionisation balance, and minimisation of the slope between A(Fe~{\sc i}) and log(EW/$\lambda$).
    \item[{\em R.1}] using {\em Gaia} {\sc edr3} and {\sc 2MASS}, we derived the bolometric absolute magnitudes, M$_{\rm bol}$, of our stars. With the inferred first set of stellar parameters ($T_{\rm eff}0$, $\log{g}_0$, [Fe/H]$_0$, $\xi_0$), we derived, through isochrone fitting, their mass, $M$. From M$_{\rm bol}$, $M$, and \teff\/, we estimate the photometric gravity, $\log{g}_{\rm phot}$, using the following equation
\begin{equation}
\log{g}_{\rm phot}= \log(M/M_{\odot})+0.4\cdot M_{bol}+4\cdot \log(T_{\rm eff})-12.505 
\end{equation}
where $M/M_{\odot}$ is the stellar mass (in solar mass units) computed with the stellar parameters from the previous run, $M_{\rm bol}$ is the bolometric magnitude (see Section~\ref{sec:mass} for calculation of the bolometric magnitude and of the stellar mass), T$_{\rm eff}$ is the spectroscopic temperature, and 12.505 is a normalisation factor to the Solar surface gravity.   
We repeated the spectral analysis, keeping the $\log{g}$ fixed to the $\log{g}_{\rm phot}$ value, obtaining a new set of stellar parameters ($T_{\rm eff}1$, $\log{g}_{\rm phot}1$, [Fe/H]$_1$, $\xi_1$). 
For some stars (29), the microturbulence thus obtained reaches its lowest minimum allowed in our minimisation procedure, i.e., 0.1 km s$^{-1}$. For this sub-sample, we repeated the spectral analysis fixing both the surface gravity and the microturbulence at its theoretical value,  as from the relation obtained for the {\em Gaia}-ESO survey for dwarf stars that we have extrapolated in few cases beyond the limits of T$_{\rm eff}$ in which it was derived. \footnote{For dwarf stars ($T_{\rm eff}\geq$ 5200 K and log~g$\geq$ 3.5): $\xi$ = 1.10 + 6.04~10$^{-4}$ (T$_{\rm eff}$-5787) + 1.45~10$^{-7}$ ($T_{\rm eff}$-5787)$^2$ - 3.33~10$^{-1}$ (log~g-4.14) + 9.77~10$^{-2}$(logg-4.14)$^2$ + 6.94~10$^{-2}$ ([Fe/H]+0.33) + 3.12  10$^{-2}$ ([Fe/H]+0.33)$^2$} 
    \item[{\em R.2}] with the stellar parameters ($T_{\rm eff}1$, $\log{g}_{\rm phot}1$, [Fe/H]$_1$, $\xi_1$) of Run 1, we recomputed the photometric gravity. We repeated the spectral analysis, keeping the $\log{g}$ fixed to $\log{g}_{\rm phot}$, obtaining the final set of stellar parameters ($T_{\rm eff}{\rm final}$, $\log{g}_{\rm phot}{\rm final}$, [Fe/H]$_{\rm final}$, $\xi_{\rm final}$). 
\end{itemize}


The catalogue with the results of the analysis is presented in the Appendix, in Table~\ref{tab:catalogue}. 
In Figure~\ref{Fig:run0-1}, we show the results of Run~0, Run~1 and final Run in the Kiel diagram for the newly observed targets and for the archival spectra. 
From the Figure, it is immediately apparent that the quality of spectro-photometric parameters has greatly improved. 
The scatter is reduced and the agreement between the two grids of isochrones (with ages from 0.1 to 14 Gyr in steps of 0.05 Gyr, at solar metallicity and at super-solar metallicity), used for a comparison, improves. This is particularly true in the area of the diagram with the lowest temperature, where the results of spectral analysis alone failed to reproduce the high gravity of the coolest stars. 
For cooler stars, however, we have a sample of stars for which the gravities are lower than those expected from the {\sc PARSEC} isochrones.

\begin{figure}
  \resizebox{\hsize}{!}{\includegraphics{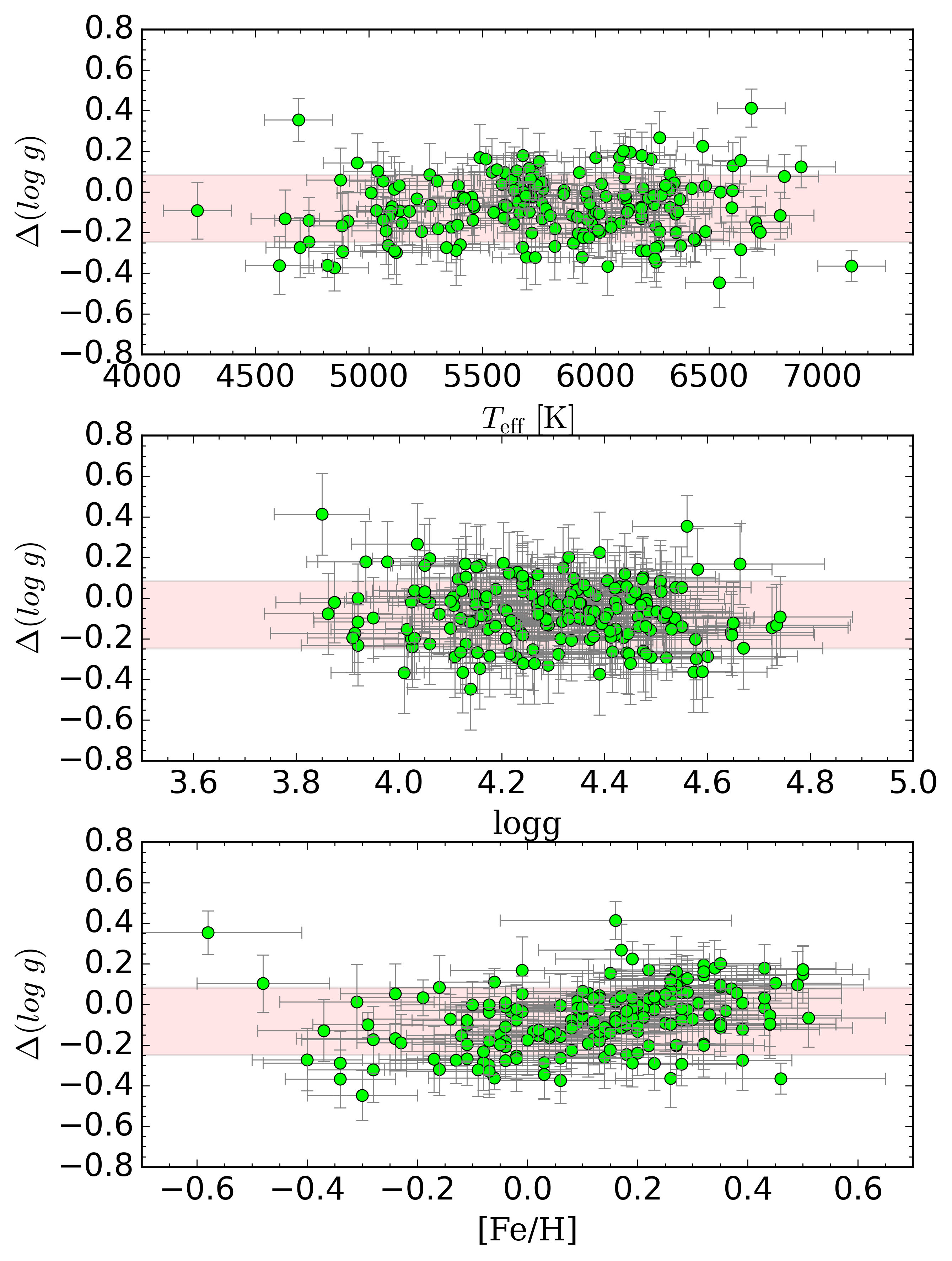}}
  \caption{$\Delta(\log{g})$ between spectroscopic and spectro-photometric gravities as a function of $T_{\rm eff}$, $\log{g}$ and [Fe/H] in the samples analysed in the present work. The shaded areas mark the average difference and its standard deviation (-0.08$\pm$0.16).    } 
  \label{Fig:delta}
\end{figure}

In Figure~\ref{Fig:delta}, we show the difference between the spectroscopic and spectro-photometric gravities ($\Delta\log{g}$) as a function of the three stellar parameters. As anticipated, the differences and scatter increase for the cooler and hotter stars in the sample, while the trends as a function of $\log{g}$ is less noticeable. There is also a slight trend of increasing $\Delta\log{g}$ with metallicity: gravities obtained from spectroscopy are slightly overestimated at high metallicity, and underestimated at low [Fe/H] with respect to the spectro-photometric ones. 
Similar results were obtained by \citet{Mortier2013}: they considered a sample of F-G-K stars and computed their surface gravities with several methods independent of spectroscopy, namely photometric transit light curve, asteroseismology and with empirical correction. 
They observed an increasing differences between spectroscopic gravity and those derived with other method with temperature. They attributed such difference to the adopted line list,  which was calibrated with solar-like stars, and it might have affected the determination of the stellar parameters of stars hotter than the Sun \citep[see, e.g.,][]{sousa11}. Their sample is limited to stars hotter than about 5000~K, thus they did not observe the opposite effect seen in the coolest K stars.  
For cooler stars, analysed in our sample, the presence of several absorption molecular bands might affect our ability to use [Fe~II] lines to constraint the surface gravity.  

The mean $\Delta(\log{g}$) is close to zero,  $\Delta(\log{g}$)=-0.08$\pm$0.16, but the scatter is  0.16, indicating that there are cases in which the spectroscopic gravity is clearly underestimated (cool stars and/or metal poor stars) and others in which it is overestimated (warm and/or metal rich stars).

\begin{figure*}
  \resizebox{\hsize}{!}{\includegraphics{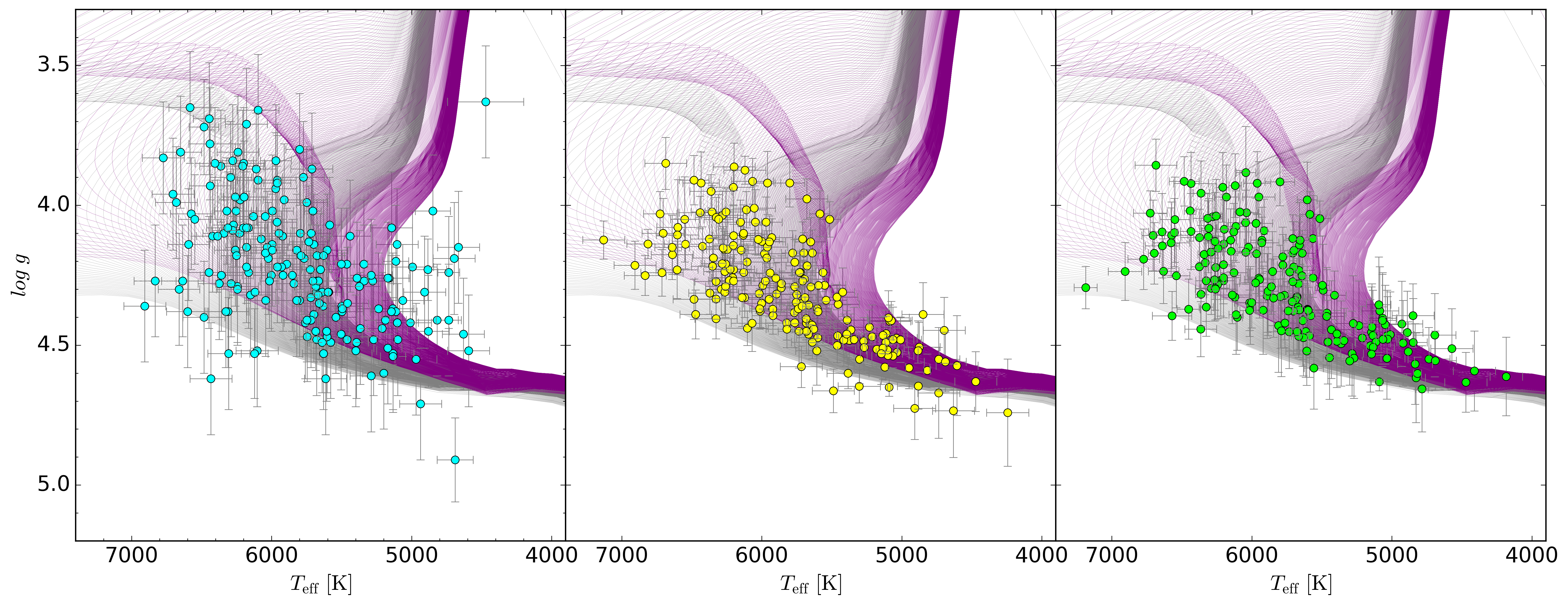}}
  \caption{Kiel diagram of the whole sample: on the  left the results from Run 0, in the central panel from Run 1, and to the right the final set of results, with $T_{\rm eff}$ from spectral analysis and $\log{g}$ from the photometry, as described in Section~\ref{sec:method}. In the three panels, we show two grids of PARSEC isochrones \citep{Bressan2012}, with ages from 0.1 to 14 Gyr, with steps of 0.05 Gyr, at solar metallicity (in grey, Z=0.013) and at super-solar metallicity (Z=0.06, in purple). } 
  \label{Fig:run0-1}
\end{figure*}


\section{Determination of stellar masses}
\label{sec:mass}

Knowledge of the properties of a planetary system is closely related to the knowledge of the parameters of its host star. The radius and mass of a planet can be derived by combining Doppler spectroscopy and transit photometry, but for that both the radius and mass of the star need to be known in advance. Moreover, the uncertainties on stellar radius and mass directly affect the uncertainties on the radius and mass of exoplanets \citep[see][for a review on the determination of stellar masses]{serenelli21}.
For the reasons stated above, it is important to perform a homogeneous determination of stellar masses, hence leading to a homogeneous set of planet properties, and ultimately providing the necessary scope for a comprehensive study of their structure and composition.

We computed stellar masses by means of a simple $\chi^2$ isochrone fitting procedure operating on the Hertzsprung--Russell (HR) diagram, i.e., by considering only the priors [Fe/H] and $T_{\rm eff}$ from our spectral analysis, and a luminosity from photometry. The underlying grid of stellar evolutionary isochrones is taken from PARSEC\footnote{\url{http://stev.oapd.inaf.it/cgi-bin/cmd}} v2.1 (the \underline{PA}dova and T\underline{R}ieste \underline{S}tellar \underline{E}volution \underline{C}ode; \citealt{Bressan2012}). Stellar masses are iteratively re-determined at each step of the analysis using the values of [Fe/H] and \teff\ derived in each run (note that $\log{g}$ is also used as input, albeit implicitly, when calculating the bolometric correction; see below for details).

The input luminosity was computed by converting the bolometric magnitude, here derived based on the $K_s$ magnitude from {\sc 2MASS} \citep{Skrutskie06}, through:
\begin{equation}
\log L= -0.4\cdot\left(K_s+BC_{K_s}-\left(-5+5\cdot\log(1/p)\right)-M_\mathrm{bol,\odot} -A_{K_s}\right),
\end{equation}
where $BC_{K_s}$ is the bolometric correction, $p$ is the parallax  in mas, $M_\mathrm{bol,\odot}$ is the bolometric magnitude of the sun ($4.8$), and $A_{K_s}$ is the absorption in the $K_s$ band. The parallax is from {\em Gaia} {\sc edr3} and its reciprocal was used as a distance proxy. The bolometric correction and absorption were calculated using the online tool YBC\footnote{\url{http://stev.oapd.inaf.it/YBC/index.html}} \citep[PARSEC Bolometric Correction;][]{Chen2019}, which interpolates a series of pre-computed bolometric correction tables in $T_\mathrm{eff}$, [Fe/H], $\log{g}$, and $E(B-V)$ using a combination of ATLAS9 \citep{Castelli2003} and Phoenix \citep{Allard2012} spectral libraries.
While $T_\mathrm{eff}$, [Fe/H], and $\log{g}$ were taken from the analysis in this work, the reddening, $E(B-V)$, was estimated using STILISM\footnote{\url{https://stilism.obspm.fr/}} \citep[STructuring by Inversion the Local InterStellar Medium;][]{Capitanio2017}.

\begin{figure}
  \resizebox{\hsize}{!}{\includegraphics{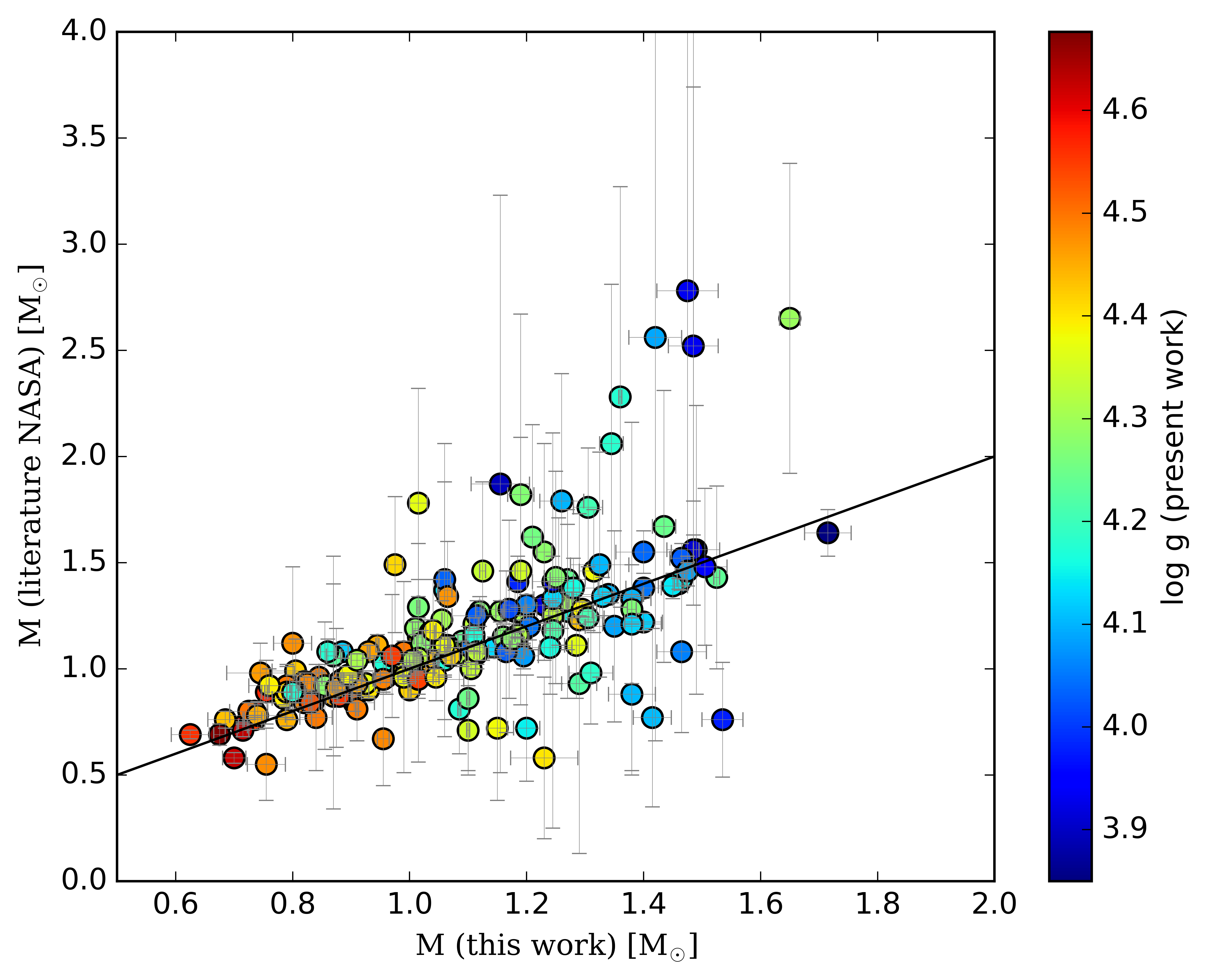}}
    \caption{Literature stellar masses (from the NASA Exoplanet Archive, or inferred values  -- see \cite{ChenKipping2017} -- when the mass was not available) as a function of the masses derived in the present work. The circles are colour-coded by the stellar surface gravity. The black line is the bisector, while the red-dashed lines are located at $\pm$0.1~M$_{\odot}$ and the blue-dashed ones at $\pm$0.2~M$_{\odot}$.  } 
  \label{fig_mass_masslit}
\end{figure}

\begin{figure}
  \resizebox{\hsize}{!}{\includegraphics{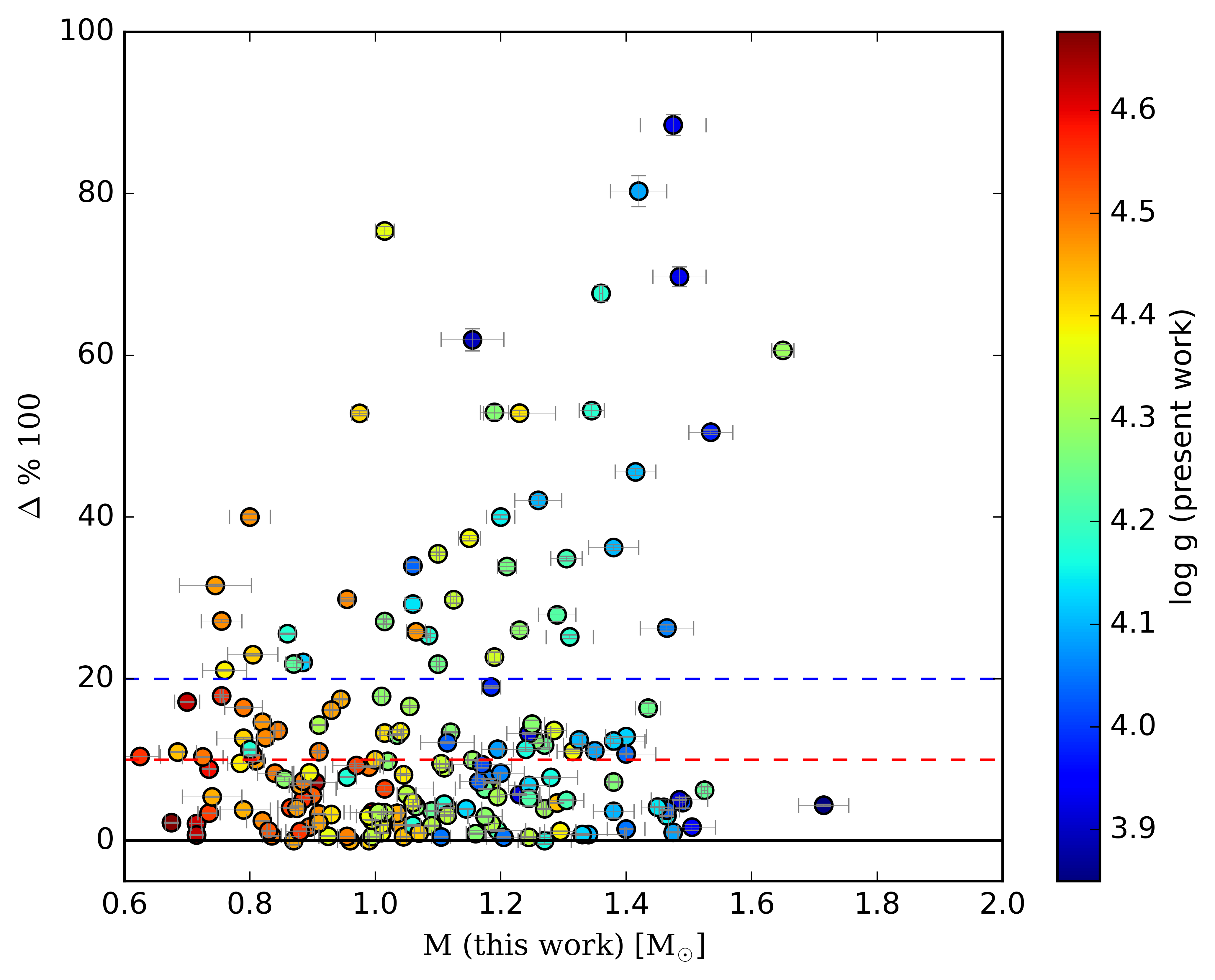}}
    \caption{Relative difference between the literature stellar masses (from the NASA catalogue, or inferred values  -- see \cite{ChenKipping2017} -- when the mass was not available) and the masses derived in the present work, as a function of the present work masses. The circles are colour-code by the stellar surface gravity. The black continuous line marks zero difference, while the red line indicates a difference within 10\%, and the blue one within 20\%.  } 
  \label{fig_mass_masslit_diff}
\end{figure}

In Figure~\ref{fig_mass_masslit}, we show the literature stellar masses as a function of the masses derived in the present work. The largest differences are for the highest masses (and lowest gravities). In addition, there is a noticeable improvement in the uncertainties, which are much lower under the current determination: the average error in the literature mass determination is 0.21$\pm$0.30, while our homogeneous mass determinations provide a mean error of 0.10$\pm$0.02. 
In Figure~\ref{fig_mass_masslit_diff}, we show the modulus of the relative differences (in percentage) versus the masses derived in the present work. For several stars with $M>1\,{\rm M}_{\odot}$, the relative differences are larger than 40\%. 
These differences justify the need for a homogeneous analysis (both in terms of the spectral analysis and mass measurement, including the grid of isochrones and the selected photometry) of the all the stars in the Ariel Reference Sample, as well as the homogeneous revaluation of the mass of their planetary companions. 

Accuracy and stellar mass precision directly reflect on the determination of the planetary mass, a parameter which plays an essential role in understanding planetary interiors, the physical processes operating in their atmospheres, as well as the planetary formation and evolutionary histories. In particular, prior knowledge of the planetary mass is important for yielding robust atmospheric properties in cloudy and secondary atmosphere exoplanets \citep{Batalha2019,Changeat2020b}. In fact, through transit spectroscopy the main degeneracies in retrieving the planetary mass are caused by the presence of clouds and the mean molecular weight \citep{Changeat2020b}. For instance, in secondary atmospheres of unknown composition the main chemical component ratio shows a direct degeneracy with the planetary mass \citep{Changeat2020b}. While planetary mass constraints less than 50\% should allow a successful retrieval analysis, a precision of 20\% is recommended for an in-depth characterisation of the atmosphere \citep{tinetti21}.


\section{Validation and quality checks}
\label{sec:validation}

\begin{figure}
  \resizebox{\hsize}{!}{\includegraphics{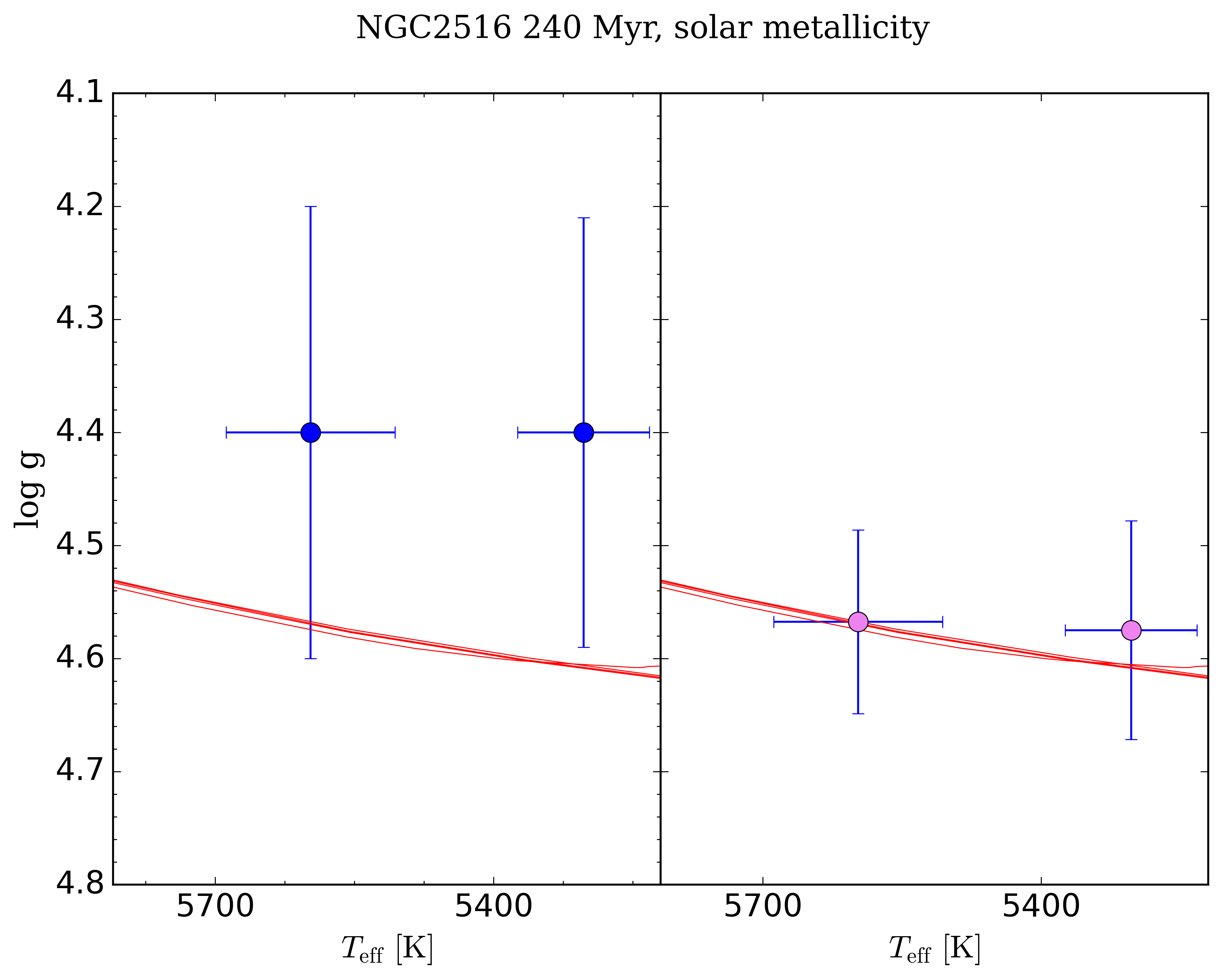}}
    \caption{Kiel diagrams of main-sequence member stars in NGC~2516 (archival spectra from the {\em Gaia}-ESO survey). In red the {\sc parsec} isochrone corresponding to the age and metallicity of NGC~2516, [Fe/H]=-0.04$\pm$0.04 from \citet{Randich2022}. On the left panel, the stellar parameters derived with a pure spectroscopic analysis, and on the right the stellar parameters from the spectro-photometric method.   } 
  \label{Fig:oc_ngc2516}
\end{figure}

\begin{figure}
  \resizebox{\hsize}{!}{\includegraphics{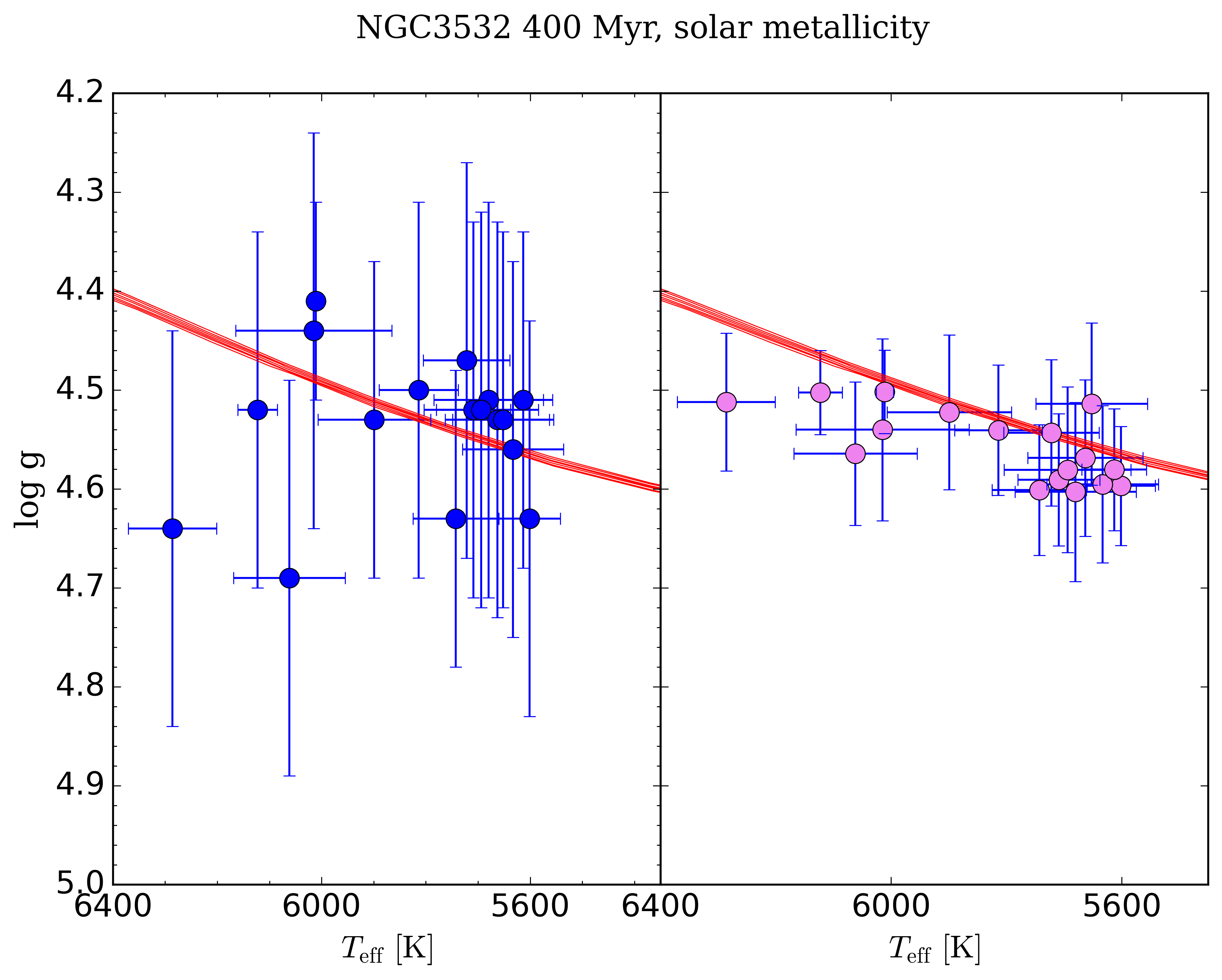}}
    \caption{Kiel diagrams of main-sequence member stars in NGC~3532, [Fe/H]=-0.03$\pm$0.08 from \citet{Randich2022}. Symbols and colours as in Figure~\ref{Fig:oc_ngc2516}.  }
  \label{Fig:oc_ngc3532}
\end{figure}

\begin{figure}
  \resizebox{\hsize}{!}{\includegraphics{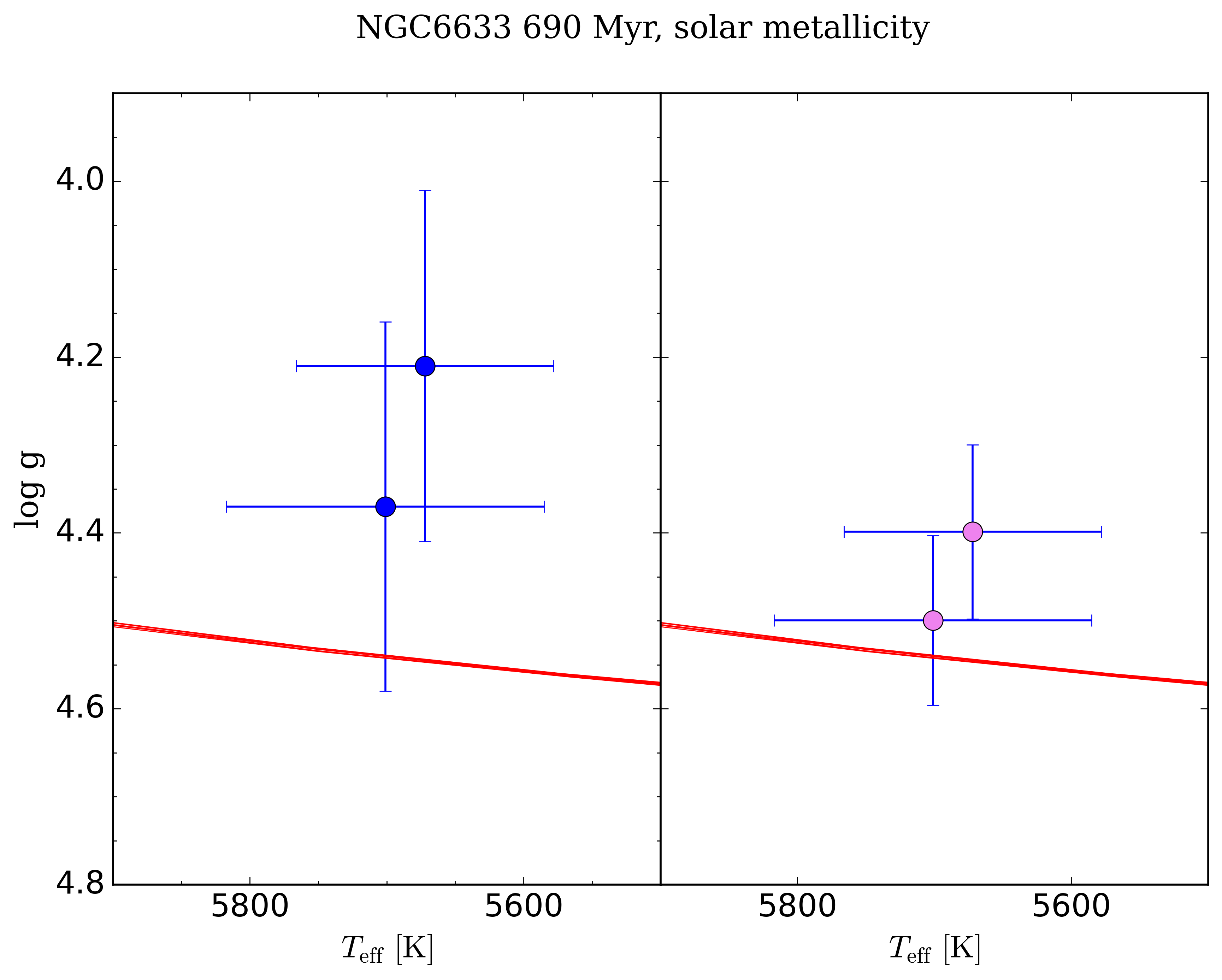}}
  \caption{Kiel diagrams of main-sequence member stars in NGC~6633, [Fe/H]=-0.03$\pm$0.04 from \citet{Randich2022}. Symbols and colours as in Figure~\ref{Fig:oc_ngc2516}.  } 
  \label{Fig:oc_ngc6633}
\end{figure}

We selected a sample of main-sequence stars, members of three open clusters that have high-resolution UVES spectra from the {\em Gaia}-ESO archival database (available from the ESO webpage). Clusters offer, indeed, the great advantage of a better determination of their ages and distances through the isochrone fitting of their sequence. For this reason they can be used as a sort of benchmark against which we can validate our method. 
The choice of clusters is not very broad, because in the {\em Gaia}-ESO sample usually the targets observed with UVES in clusters with ages > 150 Myr are giant stars, and  main sequence stars have been observed only in the case of the closest clusters. 
The three selected clusters are NGC~2516, NGC~3532 and NGC~6633, with ages ranging from  240~Myr to 690~Myr, and metallicity close to the solar one \citep{bragaglia21, Randich2022}. 
We evaluated the membership probability on the basis of the radial velocity from {\em Gaia}-ESO and of the parallax and proper motions from {\em Gaia} {\sc edr3} \citep[see][]{magrini21, jackson22}. 
We analysed the spectra of the cluster stars in a way similar to the Ariel target stars, without considering their membership to a given cluster as {\em a priori} information. 
We initially derived the stellar parameters purely spectroscopically, and then applied our iterative method to obtain the spectro-photometric parameters. 
The results are shown in Figures~\ref{Fig:oc_ngc2516}, \ref{Fig:oc_ngc3532}, and \ref{Fig:oc_ngc6633}, in which we present the stellar parameters from the spectral analysis and those derived with the spectro-photometric iterative method. The comparison with the isochrone at the age and metallicity of the clusters highlights the improvements, shown by a decrease in scatter, a reduction in the uncertainties on gravity, and better agreement with theoretical predictions. 
While the effect is most evident for the hottest stars in the validation sample, we do not have stars cooler than 5000 K to demonstrate the improvement seen in the coolest Ariel targets, for which, however, the comparison with isochrones shows the improved agreement in the Ariel sample (see Figure~\ref{Fig:run0-1}).

 We also did a sanity check on our temperature scale comparing the spectroscopic and spectro-photometric results with those obtained with the InfraRed Flux Method (hereafter IRFM), which is an almost model-independent photometric technique used to determine the T$_{\rm eff}$ of stars in a wide range of spectral types and metallicities \citep[see, e.g.][]{casagrande10, casagrande21}. 
We computed the IRFM temperatures using the intrinsic colour $(G_{\rm BP}-G_{\rm RP})_0=(G_{\rm BP}-G_{\rm RP})-E(G_{\rm BP}-G_{\rm RP})$ from {\em Gaia} {\sc DR2}, and converting it in T$_{\rm eff}$ using the relation provided by \citet{casagrande21}, in which we used our values of [Fe/H] and \logg. 
To do so we obtained the reddening $A_V$ from STILISM, and we calculated the colour excess E($G_{\rm BP}-G_{\rm RP}$) with the {\sc DR2} extinction coefficient \nicefrac{$A_V$}{$E(G_{\rm BP}-G_{\rm RP})$} by \citet{wang19}. 
The overall agreement is good with offsets of about -30~K with the temperatures of \citet{casagrande21}, and only 5\% of the sample with differences larger than $\pm$300~K.
The worst comparison is obtained for stars with the highest reddening. 
The comparison between spectroscopic and IRFM temperatures shows one of the advantages of the former ones, which is that their determination is not affected by our knowledge of the extinction. 
Finally, there are no substantial changes when comparing the IRFM temperatures and those of the first run ({\em R.0}) to the last one  ({\em R.2}). 
\section{Results}
\label{sec:results}
In what follows, we characterise our sample of Ariel Reference targets according to the properties of the planet host  stars. 

\subsection{Distribution of stellar parameters}
Target selection mainly foresees observations of main sequence stars from M to F spectral types \citep[see, e.g.,][]{Tinetti2018, Edwards2019, tinetti21}, enabling a range of combination between star and planet parameters. 
Further selection is based on the characteristics of the planetary system. Here we focus on the stellar parameters and kinematic properties of the host stars, while we refer to the paper by \cite{DelgadoMena_inprep} for their chemical characterisation. 

\begin{figure}
  \resizebox{\hsize}{!}{\includegraphics{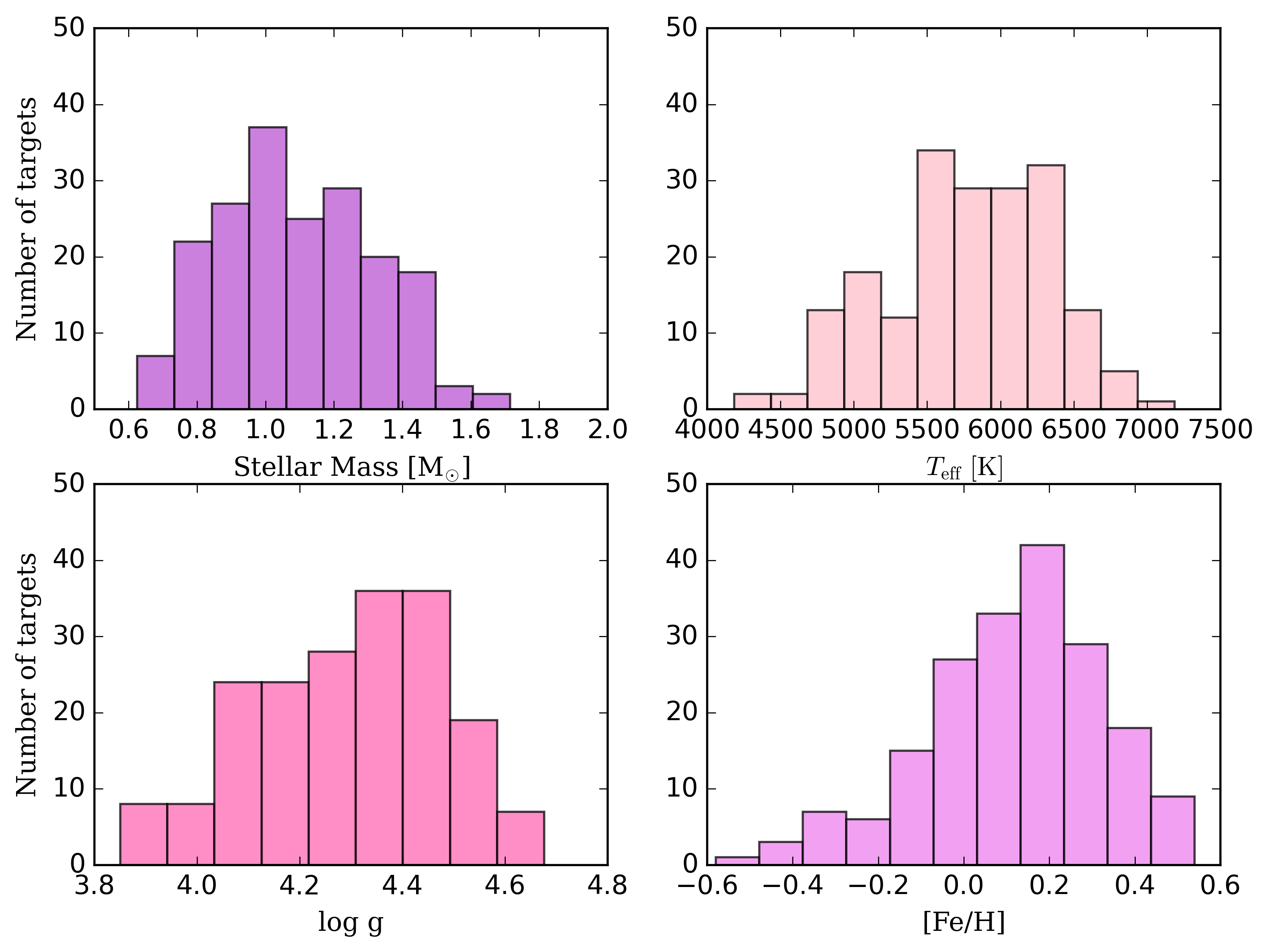}}
    \caption{Histograms of the properties of the target sample: stellar mass, T${_{\rm eff}}$, $\log{g}$, and [Fe/H].} 
  \label{Fig:histo}
\end{figure}

In Figure~\ref{Fig:histo},  histograms of mass and stellar parameter distributions are presented.
The peak of the mass distribution is between 1 and 1.3 M$_\odot$. The current sample is dominated by main sequence stars with temperatures between 5500 and 6500 K and surface gravities between 4 and 4.4, i.e., G stars. As far as metallicity is concerned, we have a clear peak at super-solar metallicity, with [Fe/H]$\sim$0.2 dex. 

\begin{figure}
  \resizebox{\hsize}{!}{\includegraphics{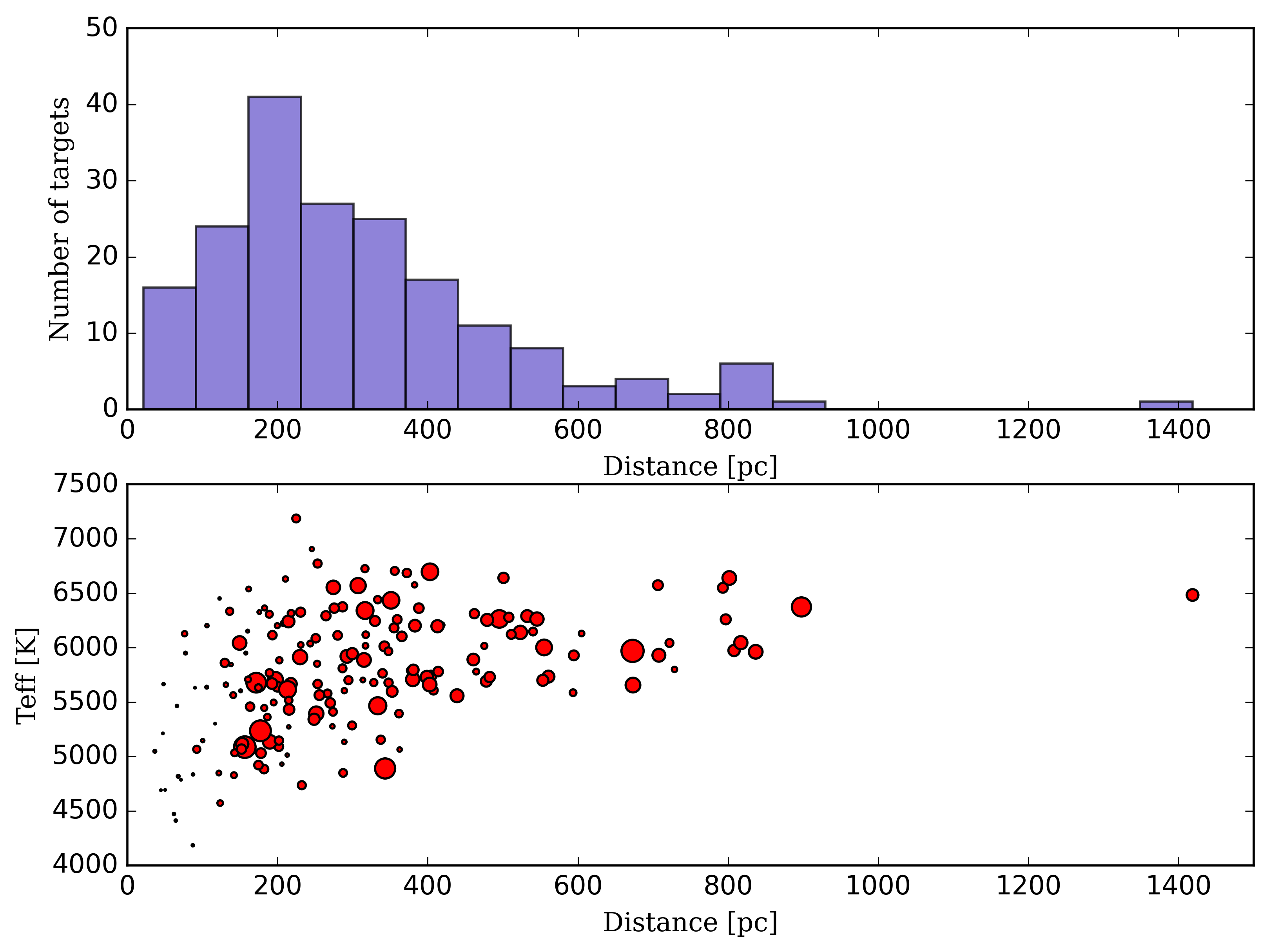}}
    \caption{Distances of the sample stars: upper panel --Histograms of the distance of the target sample from {\em Gaia} {\sc edr3}; bottom panel: \teff versus distance with the size of the circles proportional to A$_K$). } 
  \label{Fig:histo:distance}
\end{figure}

\begin{figure}
  \resizebox{\hsize}{!}{\includegraphics{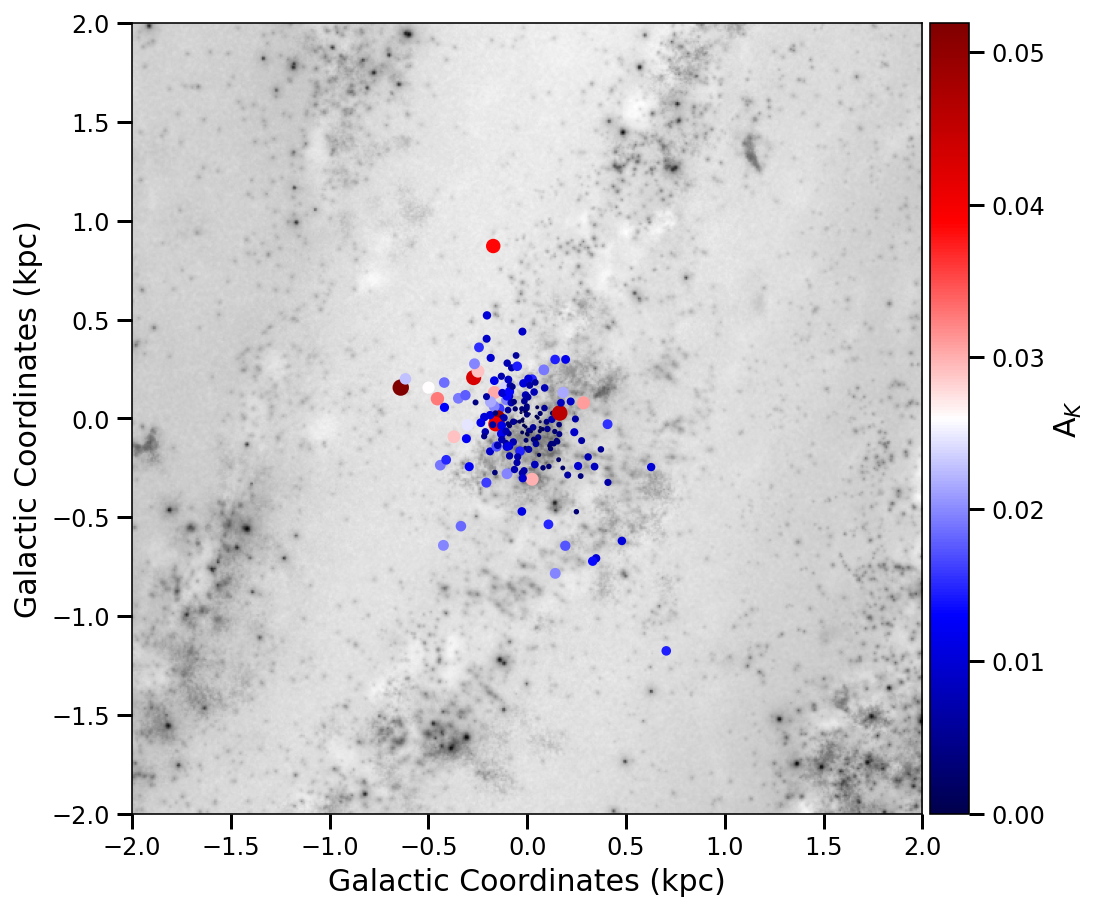}}
    \caption{Map of the sample stars, projected on the Galactic plane (colour- and size-coded by A$_K$). The region is 2.0~kpc$\times$2.0~kpc from the map of the Galaxy produced NASA/JPL-Caltech/R. Hurt (SSC/Caltech). The plot is produced with the python package {\sc mw-plot}.  } 
  \label{Fig:map}
\end{figure}

In Figure~\ref{Fig:histo:distance}, we show the distribution of distances in pc, as derived inverting the {\em Gaia} {\sc edr3} parallaxes. 
The sample is located in the solar neighbourhood, at typical distances between 200 and 400 parsec, although with a tail of stars at greater distances, up to about 1400 pc. The farthest stars are among the hottest (and thus brightest) stars of the sample.  We also show in Figure~\ref{Fig:map} the location of the sample stars, projected on the Galactic Plane. Most of them are located in the Local spiral arm. The most extinct stars are not actually the most distant ones, but those in the regions closest to the arm where gas and dust dominate.


\subsection{Orbits and kinematic properties}

The connection between the structure and evolution of the Galaxy and the ability to form and maintain planets is not yet fully understood. However, we expect that the dynamical history of stars, including stellar age and kinematics, impact the distribution and the architecture of planets in our Galaxy \citep[see, e.g.][]{michel20, bashi19, bashi20, bashi22, adi21}. 
In the era of {\em Gaia}, which is providing distances and motions of billions of stars, including planet host stars, we should take advantage of this information to relate the planetary system characteristics not only to the stellar composition, but also to its belonging to the different Galactic population  \citep[see, e.g.][for the analysis of the TESS catalogue]{carrill020}. 

Using radial velocities and proper motions from {\em Gaia} {\sc edr3}, we computed orbital parameters  of our sample stars with the {\sc galpy} package \citep{bovy15}. We assumed the model {\sc MWpotential2014} for the gravitational potential of the Milky Way. The local standard of rest (LSR) velocity was set to V$_{\rm LSR}$=220 km s$^{-1}$ \citep{bovy12} and we assumed (U, V, W)$_{\odot}$=(11.1, 12.24, 7.25) km~s$^{-1}$ for the velocity of the Sun relative to the LSR \citep{Schonrich10}. 
The orbital properties of our sample stars are presented in the Appendix, in Table~\ref{tab:orbits}. 
In Figure~\ref{Fig:toomre}, we show the location of our targets in the Toomre diagram ($\sqrt{(U^{2}+W^{2})}$, the quadratic addition of the U and W velocity components, as a function of the rotational component, V). 
In first approximation, stars with $v_{\rm pec}=\sqrt{(U^{2}+V^{2}+W^{2})}$< 50~km s$^{-1}$ belong to the thin disc, while stars with 70~km s$^{-1}$ <$v_{\rm pec}$<200~km s$^{-1}$ are part of the thick disc. 
Most of our sample stars are, as expected, part of the thin disc population, while few of them belong to the thick disc \citep[see][for a larger sample of planet host stars]{bashi22}. 
The figure also shows the eccentricity of the orbit. Stars in the thin disc, and with lower peculiar velocities, are characterised by circular orbits, with eccentricity close to zero. Stars with higher peculiar velocities have in many cases eccentricities > 0.2, i.e., more elliptical orbits and may be subject to stellar migration. It is interesting that the target sample, although limited to the solar neighbourhood, includes a variety of different populations, such as thin and thick disc. This allows us to analyse planetary systems born under different conditions from each other \citep[see][]{carrill020, bashi22} and evolved in different stellar environments. Furthermore, it will allow to study the effects on both their planetary architectures and the composition of their planets \citep{Turrini2021}.

\begin{figure}
  \resizebox{\hsize}{!}{\includegraphics{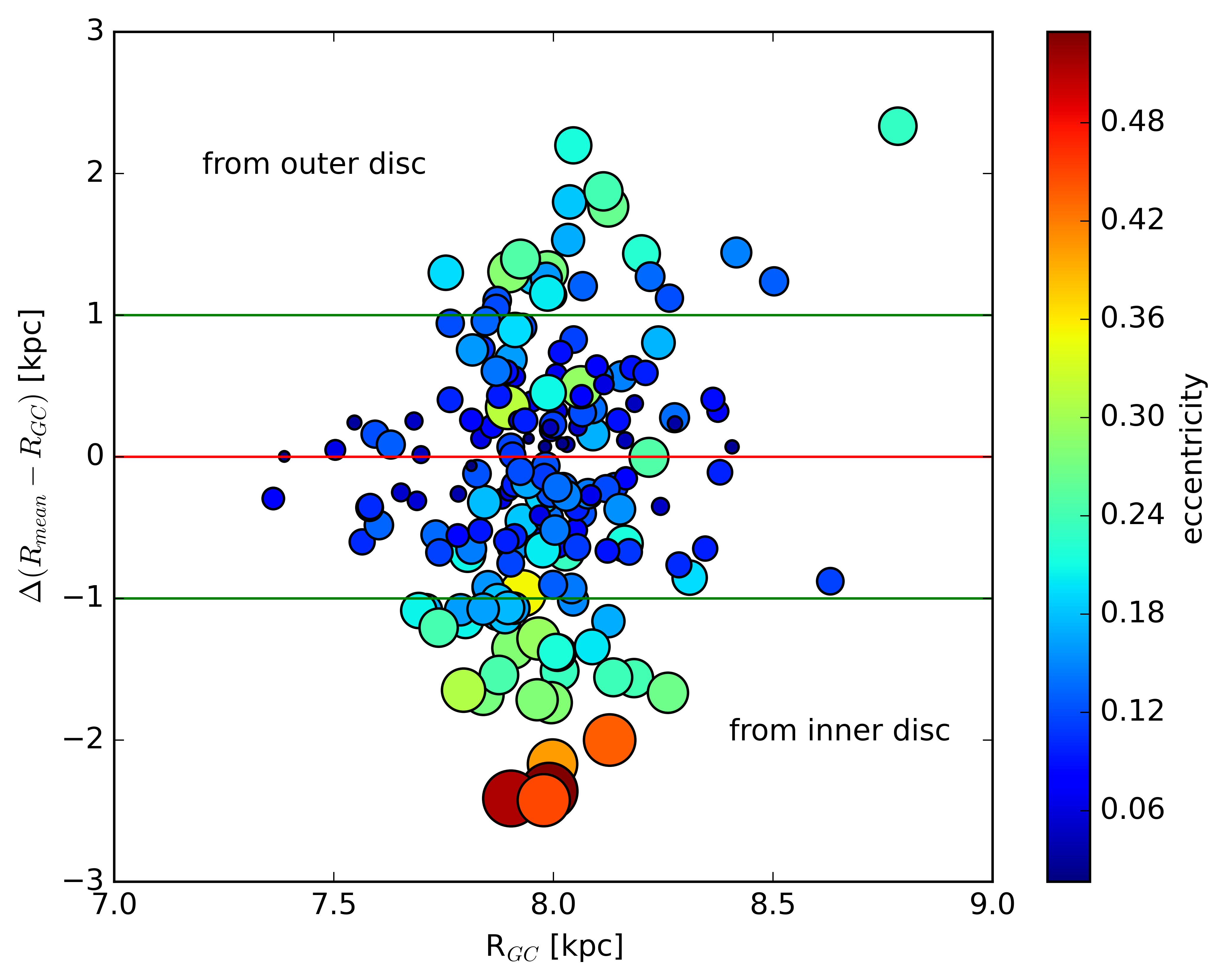}}
    \caption{Orbital radial variation ($\Delta$(R$_{mean}$-R$_{\rm GC}$)) versus Galactocentric distance (R$_{\rm GC}$) of our sample stars. Target sample stars are colour- and size-coded by the eccentricity of their orbit. The red line marks a null difference between the R$_{mean}$ of the orbit and the current R$_{\rm GC}$, while the green lines indicate a positive difference of 1~kpc (stars that likely migrated from the outer disc have $\Delta$>1~kpc) and a negative difference of -1~kpc (stars that likely migrated from the inner disc have $\Delta$<-1~kpc).   } 
  \label{Fig:migrations}
\end{figure}

Another way to show the variety of environments in which the stars that now populate the solar neighbourhood formed is by comparing their current galactocentric position with the average position of their orbits (calculated as the average of apogalactic and perigalactic radii).
In Figure~\ref{Fig:migrations}, we show the orbital radial variation ($\Delta$(R$_{mean}$-R$_{\rm GC}$)) versus Galactocentric distance (R$_{\rm GC}$) of our sample stars. 
Most of them actually have circular orbits and variations of less than 1 kpc, but there are some stars that have larger variations and probably originated in the outer or inner part of the Galactic disc. 
In this case, we are only considering one of the two causes of stellar migration: interaction with spiral arms, resonance regions or bars, which cause orbits to lose their circularity, an effect now known as ‘blurring’ \citep[see. e.g.][]{feltzing20}. The imprint of the diffusive process, ‘churning’, is usually lost when studying the orbital properties of a star, since it moves a star from circular orbit to another circular one \citep{sellwoodbinney02}.
So, in principle, we cannot exclude that all stars with circular orbits were born at the very R$_{\rm GC}$ where we have observed them now.

\begin{figure}
  \resizebox{\hsize}{!}{\includegraphics{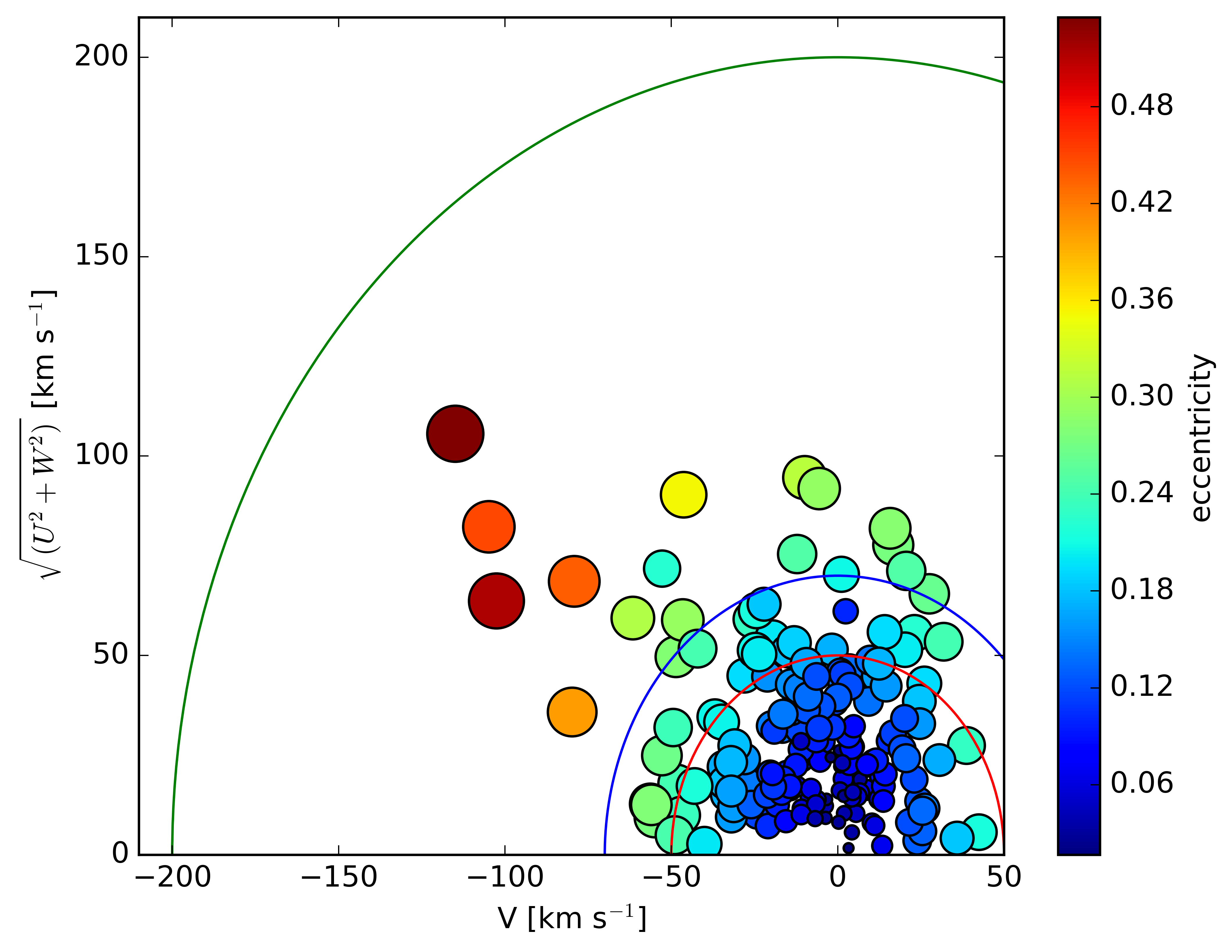}}
    \caption{Toomre diagram of our sample stars. Continuous circles indicate constant peculiar space velocities, $v_{\rm pec}=\sqrt{(U^{2}+V^{2}+W^{2})}$, at 50 (red), 70 (blue), and 200 (green) km~s$^{-1}$, as in \citet{bensby03}. 
    Target sample stars are colour- and size-coded by the eccentricity of their orbit. Stars within 50~km s$^{-1}$ can be considered with high probability to belong to the thin disc, while stars with 70~km s$^{-1}$ <$v_{\rm pec}$<200~km s$^{-1}$ are with high-probability thick disc stars. Stars with $v_{\rm pec}$ between 50 and 70~km~s$^{-1}$ can belong to both discs.  } 
  \label{Fig:toomre}
\end{figure}

\section{Relations with the characteristics of the planet}
\label{sec:relations}

Using the newly derived set of stellar parameters, we investigated the correlations between the properties of the stars in our sample and their planetary systems. We used the information on the planetary systems included in the Ariel Reference sample used for this study (we refer to \citealt{Edwards2019} for details on how the list was built up from the  NASA exoplanet catalogue\footnote{\url{https://exoplanetarchive.ipac.caltech.edu}}).
For stars with more than one planet (a total of $\sim$3\% multiplanet systems), we have included information on all their planets. 
In the following figures, the properties of planets, namely their massed and radii, are obtained from the literature, and thus are not homogeneous.  Their re-determination is currently beyond the scope of this paper, but it will be one of the issues that Ariel WGs will address in future works. On the contrary, for stellar properties we use our homogeneous values. We note that the homogeneous determination of the stellar radii for this sample of stars will be presented in an upcoming study by the ``Stellar Characterisation'' WG. Likewise to the planetary mass case, the planetary radius value hinges on the precise and accurate measurement of the stellar radius, that needs to be estimated in a coherent way to the rest of the stellar properties.
When homogeneous stellar masses and radii are known it will be possible to determine the absolute planetary masses and radii, and by that their precise planetary bulk composition. The latter, together with the atmospheric composition and atmospheric thermal structure measured by Ariel, will allow understanding both the actual distribution of heavy elements in giant planets, and the link between ice-to-rock ratios and formation mechanisms for both giant and intermediate-mass planets. We refer to \citet[][and references therein]{Helled2021} for a thorough discussion on planetary interior studies of different classes of exoplanets with Ariel.

\begin{figure}
  \resizebox{\hsize}{!}{\includegraphics{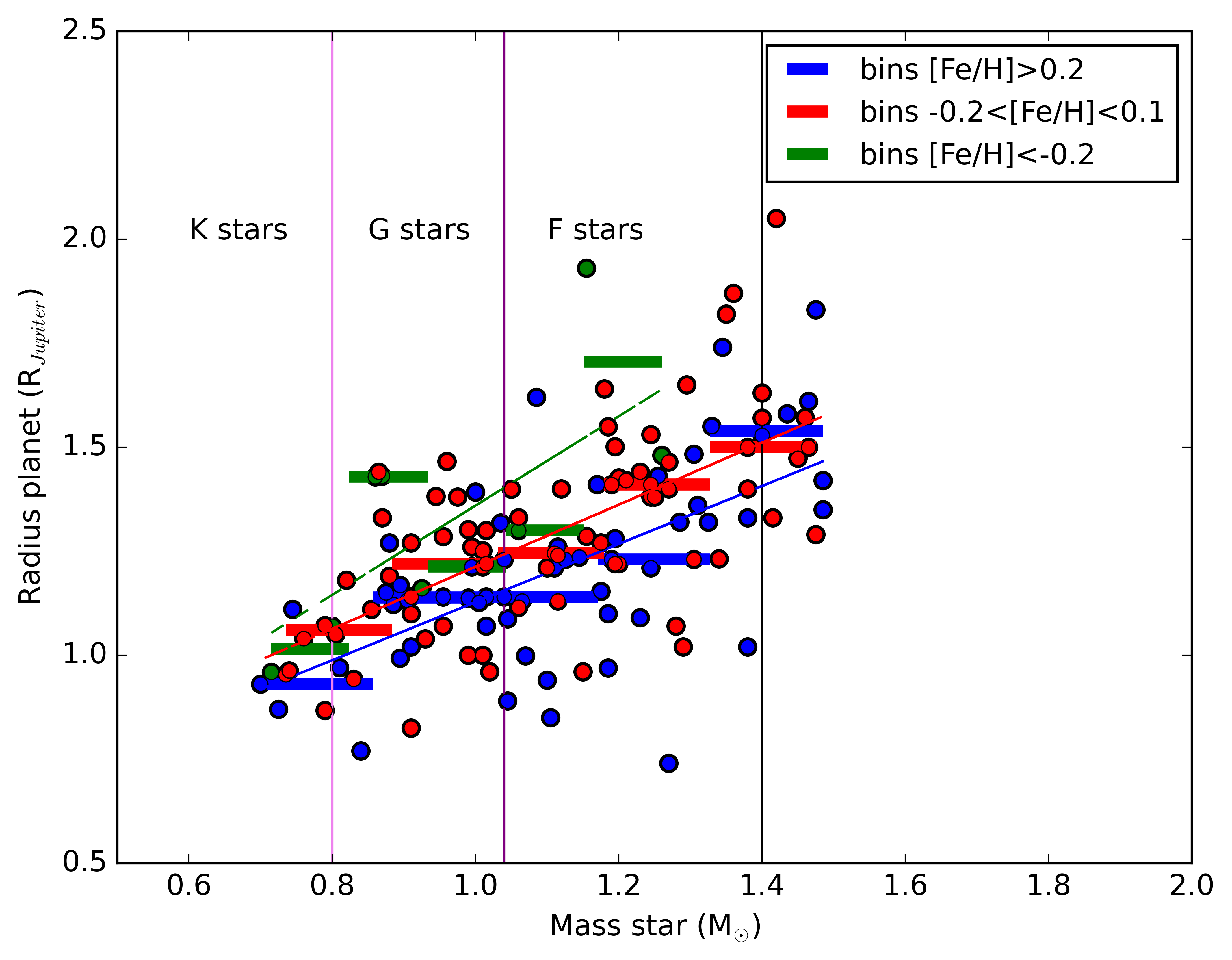}}
    \caption{Planet radius as a function of stellar mass: in blue, supersolar metallicities ([Fe/H]>+0.2), in red solar metallicities (-0.1<[Fe/H]$\leq$+0.2) and in green subsolar metallicities ([Fe/H]$\leq$-0.1). The curves are the linear fits to the three datasets, colour-coded in the same way. The horizontal lines are the median values in each binned interval. In the plot, we have excluded sub-Jovian planets (see text).
    The three vertical lines show the approximate separation among the three spectral classes. } 
  \label{Fig:radius}
\end{figure}

\begin{figure}
  \resizebox{\hsize}{!}{\includegraphics{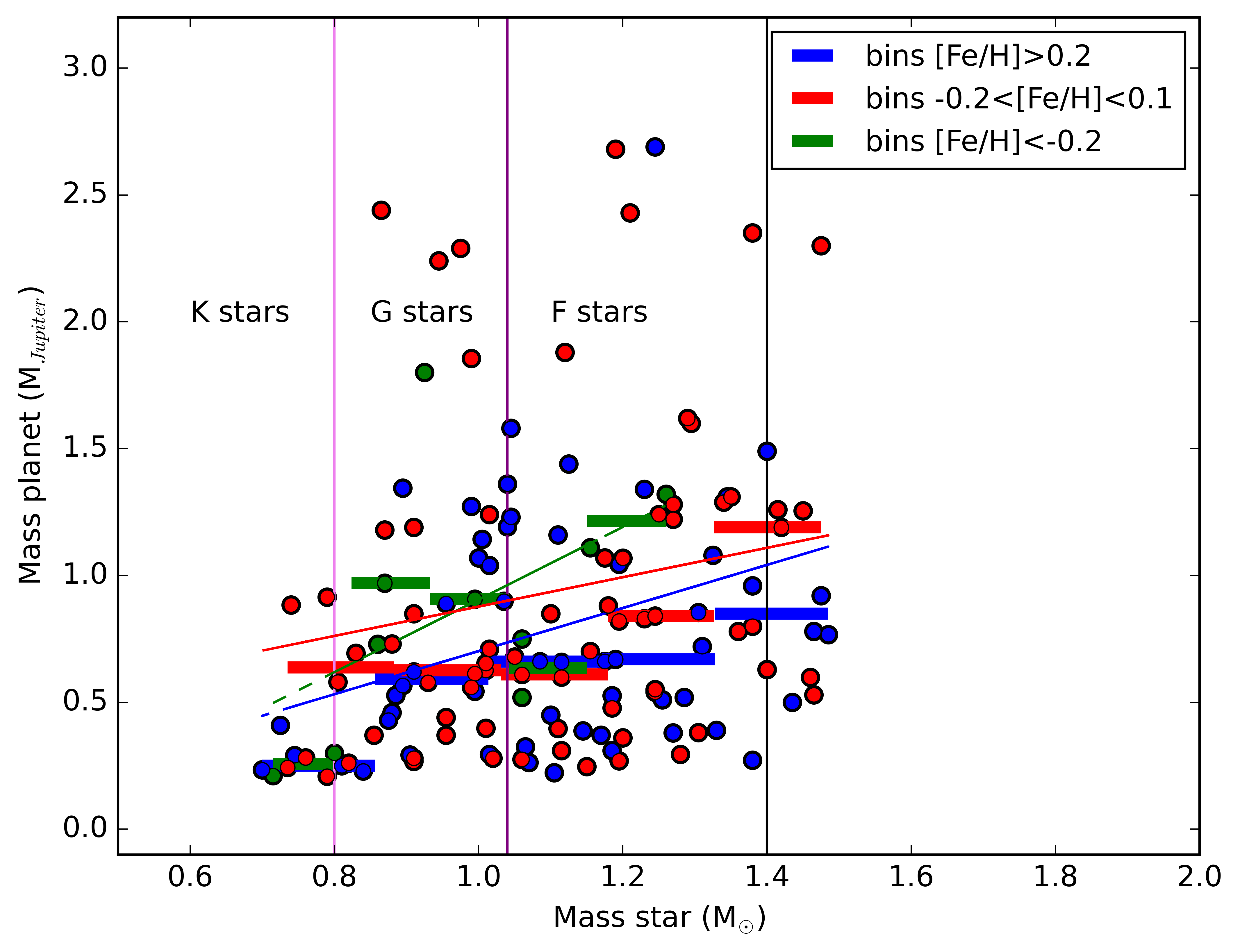}}
    \caption{Planet mass as a function of stellar mass: in blue, supersolar metallicities ([Fe/H]>+0.2), in red solar metallicities (-0.1<[Fe/H]$\leq$+0.2) and in green subsolar metallicities ([Fe/H]$\leq$-0.1). The curves are the linear fits to the three datasets, colour-coded in the same way. The horizontal lines are the median values in each binned interval. In the plot, we have excluded sub-Jovian planets (see text). The three vertical lines show the approximate separation among the three spectral classes.   } 
  \label{Fig:mass}
\end{figure}

In Figures~\ref{Fig:radius} and \ref{Fig:mass} we present the radius and the mass of the planets, respectively, plotted as a function of the stellar mass. 
Both planetary quantities show a global dependence on the stellar mass, resulting in more massive and larger planets around more massive stars. The planetary radius shows a more marked dependence on the stellar mass than the planetary mass.
%

Recent investigations have shown that planets are very diversified objects and they do not follow a single mass-radius relation. It was first suggested that small-mass planets orbiting stars with M<1 M$_{\odot}$ exhibit a relationship between the mass of the star and the radius of the planet, with larger planets around more massive stars \citep[see also][]{neil20, km21, taut22}. 
\citet{pasqucci18}, focusing on  GKM stars, showed that the radius of the planets depends on the mass of the stellar host. 
Very recently, \citet{Lozovsky21} confirmed the tendency, and explored the reasons for which more massive stars host larger planets. 
While the hypothesis bases on thermal inflation and on the effect of planet mass on the radius are less favourable, \citet{Lozovsky21} concluded that planets around more massive stars tend to be richer in gas 
and are, therefore, larger.
Specifically, they suggested that the difference in planetary radii among planets that orbit stars with different masses might be caused by different H-He mass fractions: planets around hotter and more massive stars being richer in such light elements than planets around cooler and less massive stars.

This has important implication for the mechanisms of core growth, indicating that it is more efficient both in higher-mass discs \citep{Mordasini12, venturini20} and around higher-mass stars \citep{Kennedy08}. The higher efficiency allows them to accrete a substantial gaseous envelope before the gas disc dissipates.
The results shown in Figures~\ref{Fig:radius} and  \ref{Fig:mass} confirm the finding of \citet{Lozovsky21}, extending them at higher stellar effective temperatures. For F stars, \citet{pasqucci18} expected that the presence of other effects, such as atmospheric loss by photoevaporation, might affect the properties of their planetary systems. However, in our sample the planets around F stars still follow the same relation of the planets orbiting less massive stars. 
%

To investigate the dependence on the stellar metallicity, in Figures 
 \ref{Fig:radius} and \ref{Fig:mass}  we divided the sample in three metallicity bins: a super-solar bin ([Fe/H]>+0.2), a solar bin (-0.1<[Fe/H]$\leq$+0.2) and finally a sub-solar bin ([Fe/H]$\leq$-0.1). We focused on Jovian planets (R$_{P}$ > 0.6 R$_{\rm Jup}$ and, equivalently, M$_{P}$ > 0.2 M$_{\rm Jup}$) as the statistics for smaller giant planets is too limited and the nature of these planets in our population is uncertain (i.e., small gas giants or large Neptune-like planets).
The linear fits to the data show a variation of the slope between the planetary mass, radius and the stellar mass: in Figure \ref{Fig:radius}, where such trends are more marked, the slope is flatter in the super-solar regime (0.70$\pm$0.15), and it increases with decreasing metallicity, becoming 0.75$\pm$0.11 in the solar bin, and 1.07$\pm$0.31 at lower metallicity.
Figure \ref{Fig:radius} shows that, in the size range of Jovian to super-Jovian planets, for any fixed stellar mass the planets formed around stars with super-solar metallicity possess smaller radii than their counterparts around stars with sub-solar metallicity. 
A natural explanation to this trend could be that, the larger amounts of heavy elements that can be accreted  by the giant planets during their formation and migration around higher metallicity stars \citep{Thorngren2016,Shibata2020,Turrini2021,Turrini2021b}, translates into higher densities and more compact radii (see e.g. \citealt{SuarezMascareno2021} for the illustrative case of V1298 Tau) than those of similar planets formed in lower metallicity environments. The latter coming to be comparatively richer in the light elements H and He, and hence having larger radii. \\
\indent As mentioned above, Figure \ref{Fig:mass} shows a similar yet less pronounced trend also for the planetary masses around stars with different metallicity. Planets around less metallic stars appear to reach, on average, larger masses than their counterparts around stars with solar and super-solar metallicity. Similar results have been obtained by \cite{Maldonadoetal2019} for FGKM stars in different evolutionary stage. In reality, we can see that giant planets around stars with solar and super-solar metallicity span the same range of masses as giant planets around stars with sub-solar metallicity, yet they are characterised by a larger population (relative to our sample) of sub-Jovian planets.\\
A tentative explanation for this could be that the giant planets observed around star with sub-solar metallicity are planets that formed early in the life of their circumstellar discs. Recent works have indeed demonstrated that mass accretion process seems to last longer and have higher rates in low metallicity environments when compared with solar-metallicity nearby regions (\citealt{DeMarchietal2017, Biazzoetal2019}, and references therein). The resulting higher disc mass accretion rates could then allow for forming planetary cores notwithstanding the lower abundance of solids \citep{Johansen2019} and support the runaway gas accretion phase with a larger gas supply \citep{Johansen2019,Tanaka2020}. The higher abundance of solids in the discs surrounding stars with solar and super-solar metallicity could make it possible for giant planets to form their cores over longer timescales notwithstanding the decreasing disc mass accretion rates \citep{Johansen2019}, which however would result in a lower gas supply (hence planetary mass) during their runaway gas accretion process \citep{Johansen2019,Tanaka2020}.\\
\indent Finally it is worth noting that, while the same explanation could be in principle invoked to explain also the trend of decreasing planetary radii for increasing stellar metallicity values, the comparison between Figures \ref{Fig:radius} and \ref{Fig:mass} suggests that both the core formation timescale and the planetary density play a role in shaping the observed trends. Specifically, if we focus on planets around stars of 1.2 M$_{\odot}$ we can see that the average planetary mass increases by about 50\% when going from super-solar to sub-solar metallicity values. The corresponding growth in the average radius, however, results in volumes about twice as large, meaning that the giant planets around stars with sub-solar metallicity should be less dense than their counterparts around more metallic stars. 


\section{Summary and conclusions}
\label{sec:conclusions}
We presented the spectro-photometric method developed by the Ariel mission Working Group for Stellar Characterisation of planet hosting-stars included in the Ariel reference sample. This method uses {\em Gaia} data and high-resolution, high S/N spectra to obtain stellar parameters. We validated our method, analysing member stars of three open clusters, having the same spectral types of the Ariel targets. 

We presented here the atmospheric parameters for a sample of 187 F-G-K targets in the form of catalogue, which it will be updated and provided to the Ariel Consortium and the whole scientific community on a regular basis. 
With the first catalogue we have analysed the global properties of the sample, including the distribution of the stellar parameters, their distances and orbits. We have identified a larger population belonging to the thin disc, and few stars likely belonging to the thick disc. 
Our homogeneous determination of the stellar masses highlighted a discrepancy with the stellar mass values reported in the literature. We found that 15\% in our sample has differences between 20\% and 40\%, and 5\% has differences higher than 40\%.  Furthermore, our method refined the stellar mass precision from 30\% (average of literature values) to an average of 2\%. Our work highlights the need for an homogeneous revaluation of the planetary mass too, justified by the need of a statistical approach on the study of exoplanets atmospheres to constrain planetary formation and evolution processes.
Using literature results for the characteristics of the planetary systems, in particular their radius and mass, we explored correlations between the stellar and planetary properties. We found an interesting correlation between the mass of the star and the radius of the planet, with larger planets orbiting around more massive stars and larger planets around metal poorer stars, at a given stellar mass. Giant planets orbiting sub-solar metallicity stars are probably planets that formed early in the life of their circumstellar discs. Due to the fewer amounts of heavy elements they accreted, compared to their more metal-rich counterparts, they present a lower density and hence a larger radius.
Overall we find that both core formation timescale and planetary density play a role in shaping the observed correlations.
These considerations are although pending confirmation of an homogeneous planetary radii and mass distribution determination.

\begin{acknowledgements}
The authors acknowledge the support of the ARIEL ASI-INAF agreement n.2021-5-HH.0.
The team is very grateful to the service astronomers that performed our observations at ESO (with UVES during P105, P106), at the TNG (with HARPS-N during A41, A42) and at the LBT (with PEPSI during 2021).
Based on observations collected at the European Southern Observatory under ESO programmes 105.20P2.001 and 106.21QS.001  at the LBT observatory under programme 2021\_2022\_25, and at the Italian Telescopio Nazionale Galileo (TNG) under programmes AOT41\_TAC25 and AOT42\_TAC20. We acknowledge the use of the archival data from the following ESO programs: 093.C-0417, 094.C-0428, 095.C-0367, 096.C-0417, 097.C-0571, 098.C-0292, 099.C-0374, 0101.C-0407, 099.C-0491, 0100.C-0487, 098.C-0820, 099.C-0303, 0100.C-0474, 0101.C-0497, 097.C-0948, 098.C-0860, 0100.C-0808, 0101.C-0829, 074.C-0364, 60.A-9700(G), 198.C-0169, 096.C-0657, 0102.C-0618, 0102.C-0319, 191.C-0873.
The LBT is an international collaboration among institutions in the United States, Italy and Germany. LBT Corporation partners are: Istituto Nazionale di Astrofisica, Italy; The University of Arizona on behalf of the Arizona Board of Regents; LBT Beteiligungsgesellschaft, Germany, representing the Max-Planck Society, The Leibniz Institute for Astrophysics Potsdam, and Heidelberg University; The Ohio State University, representing OSU, University of Notre Dame, University of Minnesota and University of Virginia. The TNG is operated by the Fundaci\'{o}n Galileo Galilei (FGG) of the Istituto Nazionale di Astrofisica (INAF) at the Observatorio del Roque de los Muchachos (La Palma, Canary Islands, Spain).
This research has made use of the NASA Exoplanet Archive, which is operated by the California Institute of Technology, under contract with the National Aeronautics and Space Administration under the Exoplanet Exploration Program.
C.D. acknowledges financial support from the State Agency for Research of the Spanish MCIU through the Center of Excellence Severo Ochoa award to the Instituto de Astrofísica de Andalucía (SEV-2017-0709) and the Group project Ref. PID2019-110689RB-I00/AEI/10.13039/501100011033. This work was supported by Funda\c c\~ao para a Ci\^encia e a Tecnologia (FCT) through the research grants UIDB/04434/2020 and UIDP/04434/2020. D.B. is supported in the form of work contract FCT/MCTES through national funds and by FEDER through COMPETE2020 in connection to these grants: UID/FIS/04434/2019; PTDC/FIS-AST/30389/2017 \& POCI-01-0145-FEDER-030389. T.L.C.~is supported by FCT in the form of a work contract (CEECIND/00476/2018).
N.S. and M.T.  acknowledges  the  financial  support  to  this  research by INAF, through the Mainstream Grant 1.05.01.86.22 assigned to the project “Chemo-dynamics of globular clusters: the Gaia revolution” (P.I. E. Pancino). E.D.M., V.A. and S.G.S. acknowledge the support from Funda\c{c}\~ao para a Ci\^encia e a Tecnologia (FCT) through national funds
and by FEDER through COMPETE2020 by the following grants PTDC/FIS-AST/28953/2017, POCI-01-0145-FEDER-028953 \& PTDC/FIS-AST/32113/2017, POCI-01-0145-FEDER-032113. E.D.M. acknowledges the support by the Investigador FCT contract IF/00849/2015/CP1273/CT0003 and in the form of an exploratory project with the same reference. V.A. and S.G.S. also acknowledge the support from FCT through the contracts IF/00650/2015/CP1273/CT0001 and CEECIND/00826/2018 funded by FCT (Portugal) and POPH/FSE (EC). M.R. acknowledges support from the Italian Space Agency (ASI) under contract 2018-24-HH.0.
G.B. acknowledges support from CHEOPS ASI-INAF agreement n. 2019-29-HH.0.
D.T. acknowledges the support of the Italian National Institute of Astrophysics (INAF) through the INAF Main Stream project "Ariel and the astrochemical link between circumstellar discs and planets" (CUP: C54I19000700005). 
M.T., L.M., MVdS, A.B. acknowledge the following grants: MIUR Premiale "{\em Gaia}-ESO survey" (PI S. Randich), MIUR Premiale "MiTiC: Mining the Cosmos" (PI B. Garilli). 
M.T. acknowledges the ASI-INAF contract 2014-049-R.O: "Realizzazione attivit\`a tecniche/scientifiche presso ASDC" (PI A. Antonelli), Fondazione Cassa di Risparmio di Firenze, progetto: "Know the star, know the planet" (PI E. Pancino). 
\end{acknowledgements}

\bibliographystyle{aa}
\bibliography{Biblio}


\begin{appendix}
\section{Additional material}

In this section, we provide Table~\ref{tab:catalogue} with the catalogue, in which we give the star ID, the stellar mass (M), and its upper (M$_U$) and lower (M$_L$) limit, the effective temperature (\teff), surface gravity ($\log{g}$), metallicity ([Fe/H]), microturbulent velocity ($\xi$) and their uncertainties. For [Fe/H] we provide both the uncertainties due to the scatter of the abundances derived from different Fe~I lines (e$_{\rm [Fe/H] l}$), and due to the errors on the parameters (e$_{\rm [Fe/H] p}$). 
Finally, we provide a flag for the microturbulent velocity: 0 means that $\xi$ is derived from spectral analysis, and 1 that it is assumed from relation between $\xi$ and T$_{\rm eff}$, $\log{g}$ and [Fe/H].
In the last column, S,  we give the provenance of the spectrum from which the parameters were obtained. 

In Table~\ref{tab:orbits} we provide the orbital properties of our sample of stars (with the exception of Qatar~8): the three components of the velocities (U, V, W), the Galactocentric radius (R$_{\rm GC}$) and the apogalactic and perigactic radii (R$_{apo}$ and R$_{peri}$), the eccentricity (e),  the maximum height above the Plane (Z$_{max}$) and the distance (computed inverting the {\em Gaia} {\sc dr3} parallax).

\end{appendix}

\end{document}